\documentclass{nature}
 
\usepackage[version=3]{mhchem} 
\usepackage[T1]{fontenc}       
\usepackage{color}
\usepackage{textcomp}
\usepackage{amsmath}
\usepackage{amssymb}
\usepackage[draft]{changes}
\definechangesauthor[color=orange]{DW}
\usepackage{graphicx}
\usepackage{amsmath}
\usepackage{amssymb}
\usepackage{amsfonts}
\usepackage{textcomp}
\usepackage{bm}
\usepackage{multirow}
\usepackage{float}
\usepackage{placeins}

\usepackage{color}

\usepackage{cancel}

\usepackage[font=footnotesize,labelfont=bf]{caption}
\usepackage{geometry}
 \usepackage[sort, numbers]{natbib}
 \setcitestyle{square}

\usepackage{graphicx}
\makeatletter
\let\saved@includegraphics\includegraphics
\AtBeginDocument{\let\includegraphics\saved@includegraphics}
\renewenvironment*{figure}{\@float{figure}}{\end@float}
\makeatother

\newcommand{\E}{\ensuremath{{\cal E}}}

\newcommand{\bra}[1]{\left< #1\right|}
\newcommand{\ket}[1]{\left| #1\right>}
\newcommand{\proj}[2]{\ket{#1}\!\bra{#2}}

\title{Controlled Coherent Coupling in a Quantum Dot Molecule\\ Revealed by Ultrafast Four-Wave Mixing Spectroscopy}

\author{\small
Daniel~Wigger,$^{1,2,\ast}$ Johannes~Schall,$^{3}$ Marielle~Deconinck,$^{3}$ Nikolai~Bart,$^{4}$ Pawe\l{}~Mrowi{\'n}ski,$^{1,5}$\\ Mateusz~Krzykowski,$^{1}$ Krzysztof~Gawarecki,$^{1}$ Martin~von~Helversen,$^{3}$ Ronny~Schmidt,$^{3}$ Lucas~Bremer,$^{3}$\\ Frederik~Bopp,$^{6}$ Dirk~Reuter,$^{7}$ Andreas~D.~Wieck,$^{4}$ Sven~Rodt,$^{3}$ Julien~Renard,$^{8}$ Gilles~Nogues,$^{8}$\\ Arne~Ludwig,$^{4}$ Pawe\l{}~Machnikowski,$^{1}$ Jonathan~J.~Finley,$^{6}$ Stephan~Reitzenstein,$^{3}$ Jacek~Kasprzak$^{6,8,\ast}$}

\begin{document}

\maketitle

\begin{affiliations}
\item {\small Institute of Theoretical Physics, Wroc\l{}aw University of Science and Technology, 50-370~Wroc\l{}aw, Poland}
\item {\small School of Physics, Trinity College Dublin, Dublin 2, D02 PN40, Ireland}
\item {\small Institute of Solid State Physics, Technische Universit\"at Berlin, 10623 Berlin, Germany}
\item {\small Lehrstuhl f\"ur Angewandte Festk\"orperphysik Ruhr-Universit\"at Bochum, 44780 Bochum, Germany}
\item {\small Laboratory for Optical Spectroscopy of Nanostructures, Department of Experimental Physics, Wroc\l{}aw University of Technology, 50-370 Wroc\l{}aw, Poland}
\item {\small Walter Schottky Institut and Physik Department, Technische Universit\"at M\"unchen, 85748 Garching, Germany}
\item {\small Department Physik, Universit\"at Paderborn, 33098 Paderborn, Germany}
\item {\small Universit\'e Grenoble Alpes, CNRS, Grenoble INP, Institut N\'{e}el, 38000 Grenoble, France}
\item[$^\ast$] {\small daniel.wigger@tcd.ie, jacek.kasprzak@neel.cnrs.fr}
\end{affiliations}

\begin{abstract}
\small 
Semiconductor quantum dot molecules are considered as promising candidates for quantum technological applications due to their wide tunability of optical properties and coverage of different energy scales associated with charge and spin physics. While previous works have studied the tunnel-coupling of the different excitonic charge complexes shared by the two quantum dots by conventional optical spectroscopy, we here report on the first demonstration of a coherently controlled inter-dot tunnel-coupling focusing on the quantum coherence of the optically active trion transitions. We employ ultrafast four-wave mixing spectroscopy to resonantly generate a quantum coherence in one trion complex, transfer it to and probe it in another trion configuration. With the help of theoretical modelling on different levels of complexity we give an instructive explanation of the underlying coupling mechanism and dynamical processes.\\
\begin{center}
\includegraphics[width=0.5\textwidth]{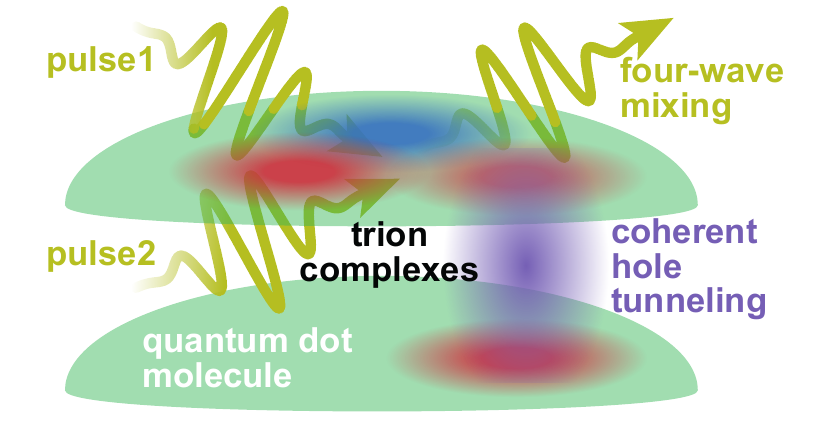}
\end{center}
\end{abstract}

\twocolumn
\footnotesize

\section{Introduction}
Maintaining and controlling coherence of quantum states spanning multiple physically distinct systems is at the core of quantum technologies, like quantum information processing and\linebreak communication~\cite{horodecki2009} or quantum metrology~\cite{giovannetti2011}. In the competitive and multi-disciplinary quest for systems that might allow one to efficiently induce, control, and detect quantum coherence, exciton complexes in semiconductor quantum dots (QDs) were already very early recognized as promising building\linebreak blocks~\cite{henneberger2016semiconductor, michler2017quantum}. QDs  offer an efficient light-matter interface~\cite{garcia2021semiconductor}, which makes it possible to control their quantum states  at sub-picosecond timescales~\cite{FrasNatPhot16}, beyond cryogenic temperatures up to ambient conditions~\cite{HolmesNanoLet14}. \\
Closely stacked QDs forming quantum dot molecules (QDMs) offer a large flexibility to control their optical and spin properties by external electric or magnetic fields~\cite{bayer2001coupling, KrennerPRL05,doty2006electrically,  doty2008optical, kim2008optical,ardelt2016coulomb}. Therefore, they are particularly attractive for quantum applications, such as quantum repeaters, requiring efficient spin-photon interfaces. While spectral characteristics of the coupling have been revealed in QDM systems by linear spectroscopy methods~\cite{bayer2001coupling, KrennerPRL05,doty2006electrically, StinaffScience06, doty2008optical, kim2008optical,ardelt2016coulomb}, these experiments are only sensitive to the structure of eigenstates and their occupations, thus yielding a typical spectral picture that essentially characterizes two coupled modes of any kind, be it classical or quantum. In contrast, tracing coherences, which are fingerprints of quantum superpositions, requires the application of a coherent nonlinear spectroscopy tool, like four-wave mixing (FWM). This poses a major challenge, in particular in the case of single quantum systems, where the optical signal is very weak. For this reason, detecting coherent characteristics of the coupling between different excitonic complexes in self-assembled QDs is still in its infancy.\\
Fortunately, the epitaxial growth and nanoprocessing of QD-based quantum devices has been improved close to perfection and individual nanostructures can nowadays be embedded deterministically into advanced nanophotonic devices with high photon extraction efficiency to improve the application potential and to enable experiments relying on high single-photon flux~\cite{rodt2020deterministically,SchallAQT21}. On the other hand, the development of heterodyning and interferometric techniques in FWM spectroscopy made it possible to apply this method to single quantum emitters and to detect and characterize the coupling between individual transitions in various systems~\cite{LangbeinOL06, MartinPRB18, KasprzakNPho11, raymer2013entangled, MermillodPRL16,MermillodOptica16, DelmontePRB17}.\\
Here, we go significantly beyond experiments on basic single structures by using QDMs with a much richer excitonic level scheme that can be controlled by an external bias. This allows us to demonstrate that the hole-tunnelling in the QDM can be used to transfer resonantly created quantum coherence between different trion transitions. Similar experiments were previously performed on quantum wells~\cite{davis2011three, langer2012magnetic,nardin2014coherent,salewski2017high}. In contrast to those, a single QDM is an atomic-like system, which offers broader perspectives for quantum technologies, e.g., interfacing spins and photons, at the same time making the experiment much more challenging. This way we demonstrate the coherent control of coherences in a coupled QD system, as required for quantum applications~\cite{FrasNatPhot16}. We further show how this coherent coupling can be controlled by applying an electric field. In quantum optical terms, this amounts to parametrically switching between a V-system and a $\Lambda$-system in a single physical structure, which allows us to address different pairs of coherences.\\
Through this first demonstration we make a crucial step toward the implementation of the ultrafast FWM methodology in photonic quantum technologies. In this way our work paves the way toward the realization of controlled long-range coherent coupling between distant solid state qubits.

\section{Device and experimental method}

\begin{figure}[t!]
    \centering
    \includegraphics[width=0.9\columnwidth]{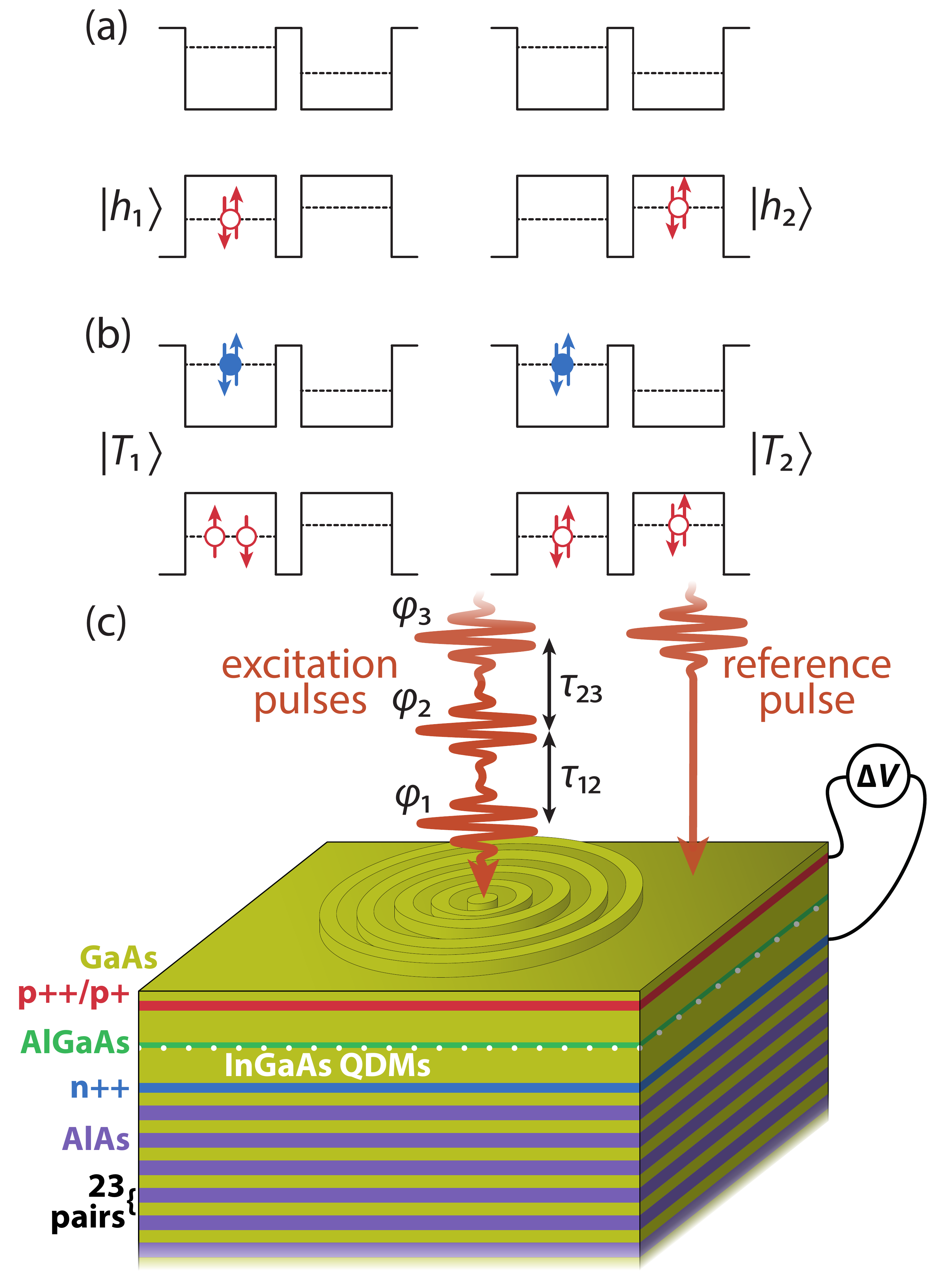}
   \caption{Involved QDM states and experimental setup. (a) Ground states of the involved trions possessing one hole in either of the quantum dots. (b) Relevant excited states formed by an additional exciton in the left quantum dot. (c) Schematic picture of the sample structure. Above a GaAs/AlAs distributed Bragg reflector the InGaAs quantum dot molecule (QDM) layer is sandwiched between charge-doped layers, which are connected to a bias source. The optical in- and out-coupling is improved by a circular Bragg grating, also called a bulls-eye photonic structure, on the top surface. The laser pulses used for the optical excitation in the four-wave mixing experiment are focused on the center of the grating, while the reference beam hits the free surface in the vicinity.}
    \label{fig:setup}
\end{figure}%

We here use the aforementioned advantages of semiconductor quantum photonic technology to demonstrate controlled coherent coupling between excitonic transitions hosted by a pair of InAs QDs separated by $\approx 8$~nm, forming a QDM~\cite{bayer2001coupling, KrennerPRL05, StinaffScience06, SchallAQT21}. The system is doped, so that, apart from spin degeneracy, the ground state manifold of the QDM consists of two states, with a hole in one or the other QD, as schematically depicted in Fig.~\ref{fig:setup}(a). As seen in the sketch in Fig.~\ref{fig:setup}(b), from the two coupled charged three-particle complexes (trions) that can be excited optically, one is entirely located in one of the QDs, while the other spans both dots. We prove the coupling between the optically induced coherences between these hole and trion states in the coupled QDM system by performing two-dimensional four-wave mixing (2D FWM) spectroscopy (see below), in which coherent coupling between different optically active transitions is revealed as off-diagonal peaks in the 2D spectra~\cite{DelmontePRB17}. Similar results can be expected for exciton transitions in the other QD, which are typically separated by several meV and are therefore not accessible in this study due to the limited spectral width of the excitation~\cite{ardelt2016coulomb}.\\
In our sample, the QDMs are placed in the structure schematically shown in Fig.~\ref{fig:setup}(c), the layer structure of this QDM quantum device supports storage of holes by an AlGaAs tunneling barrier between the QDMs and the p-doped region of the diode. To achieve suitably high optical in- and out-coupling efficiency for the FWM signal~\cite{MermillodPRL16}, a circular Bragg grating is deterministically positioned around the QDM in the center of the quantum device.  For details on the sample design and processing we refer to Ref.~\cite{SchallAQT21}. The charge-doped layers surrounding the QDM sheet provide a p-i-n diode structure allowing voltage control. While scanning the external bias voltage, we monitor the formation of off-diagonal peaks in 2D FWM spectra of the molecule.

\section{Results}

\subsection{Photoluminescence spectra}
\begin{figure}[b!]
    \centering
    \includegraphics[width=\columnwidth]{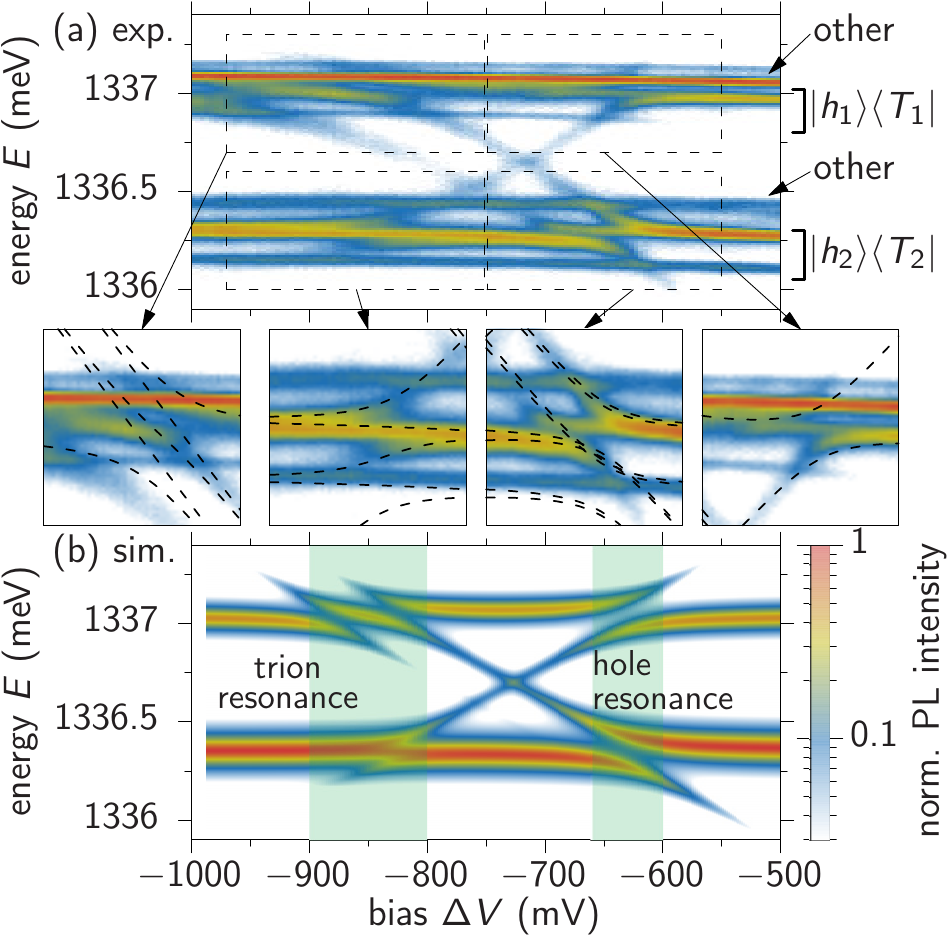}
   \caption{Bias scan of the PL spectra of the QDM-device. (a)~Measured PL spectra as a function of the applied bias. The labels on the right mark the respective trions correspond to Fig.~\ref{fig:scheme}(a) and transitions from other exciton complexes, likely neutral excitons. The zoom-ins show the calculated optical transitions as dashed lines. (b) Semi-empirical simulation of the PL spectrum. The avoided crossings marked by the green shaded areas on the left stem from trion (excited state) resonances, those on the right from hole (ground state) resonances.}
    \label{fig:PL}
\end{figure}%
~

\noindent
For characterising the QDM's optical properties we measure photoluminescence (PL) spectra for varying applied electric\linebreak fields, which leads to an energy shift of the involved electronic states with respect to each other. Consequently, the hole and trion states in the two dots are brought into resonance for specific electric fields and the tunnel-coupling leads to characteristic avoided crossings in the optical spectra. The results of this pre-characterisation are depicted in Fig.~\ref{fig:PL}(a) and are in excellent agreement with the findings in Ref.~\cite{SchallAQT21}. To further check the integrity of the emitter structure we calculate the transition energies for a heuristic, semi-empirical model of the QDM as shown by the dashed lines (details are given in the Supporting Information (SI)). We use the same fitting parameters as in Ref.~\cite{SchallAQT21} and achieve the same high level of agreement with the measured spectral lines. Note, that not all possible transitions from the rich spin configurations including singlet and triplet states predicted by the theory (dashed lines) appear as bright spectral lines in the measurement because their dipole matrix elements are small and the optical signal is not visible at the given signal-to-noise ratio. While the entire structure of the PL spectra is quite complex with a total of ten visible lines, we are particularly interested in the four pronounced avoided crossings, which indicate state hybridisations and coherent tunnel-couplings. Two avoided crossings appear around an applied bias of $\Delta V\approx -650$~mV and are attributed to the resonance between the two hole states (ground states). The other two at $\Delta V\approx -850$~mV stem from the resonance of the trion states (excited states). We also find other bright spectral lines that are unaffected by the applied electric field and do not participate in the avoided crossings (marked by 'other'). These transitions likely belong to neutral excitons in the QDM (SI).\\
In Fig.~\ref{fig:PL}(b) we show simulated PL spectra derived from a semi-empirical model~\cite{Ponomarev2006,Scheibner2007} (SI) leading to the dashed lines in the zoom-ins in (a). Due to the optical selection rules we find that not all transitions lead to bright emission lines in the spectrum (SI). The brightest lines, which form avoided crossings, are well reproduced by the simulation. In order to relate the spectra to the morphological characteristics of the QDM and to confirm the form of the spectrum with a model based on realistic wave functions and an extended number of hole and trion states, we have performed $\bm{k}\cdot \bm{p}$ calculations~\cite{bahder90,Winkler2003,pryor98b,bester06,bester06b,Caro2015,Gawarecki2018a,Krzykowski2020, Swiderski2016,Azizi2015,Karwat2021,Gawelczyk2017} of the trion PL spectra. In spite of the multiple transitions present in the $\bm{k}\cdot \bm{p}$ model, the low-excitation PL spectrum is very similar to that obtained in the semi-empirical model (SI).\\
Although the line splittings in the PL spectra certainly stem from the coupling between the two different exciton complexes, the appearance of an avoided crossing can be described already by a classical coupled oscillator model. Therefore, our goal is to directly address quantum coherence properties of the coupling mechanism via FWM micro-spectroscopy. In this experiment the QDM is resonantly excited with three laser pulses $\E_{1,2,3}$ (see Fig.~\ref{fig:setup}(c) and SI). Each of the pulses is phase-labeled by $\E_{1,2,3}\sim e^{i\varphi_{1,2,3}}$, which can experimentally be achieved by different propagation directions ${\bm k}_{1,2,3}$~\cite{shah2013ultrafast} or -- as in our case -- by radio frequency-shifts ($\Omega_1, \Omega_2, \Omega_3)=(80, 79, 79.77)$~MHz \cite{LangbeinOL06}. The considered FWM signal is then carried by the phase combination $\varphi_3+\varphi_2-\varphi_1$~\cite{wigger2018rabi,hahn2021influence}, which means that, at small pulse areas, we are detecting the third-order nonlinearity $\sim\E_1^\ast\E^{ }_2\E^{ }_3$ ($\chi^{(3)}$ regime). This FWM signal from the sample is retrieved via a heterodyne detection with a reference pulse that is focused next to the grating structure on the sample surface (Fig.~\ref{fig:setup}(c)). As explained in more detail below, the first pulse resonantly creates trion quantum coherences, which are transferred to the other trion states by tunnel-coupling and the second and third pulse. The FWM signal is then emitted with respectively different energies. By varying the delay between the first and second pulse $\tau_{12}$ and keeping $\tau_{23}=0$ fixed, we are able to probe the coherence dynamics in the system (SI). Note, that this method directly probes the quantum coherence between the involved optically active states and therefore immediately demonstrates the quantum nature of the observed avoided crossing. Besides the coherent coupling, this spectroscopy method goes well beyond conventional (\textmu PL) spetroscopy and can give valuable insight in the homogeneous broadening~\cite{HahnAdvSci21, kasprzak2022coherent}, phonon-induced dephasing processes~\cite{wigger2020acoustic}, or Rabi oscillations~\cite{wigger2018rabi}. Moreover, by changing the pulse powers we can characterise the light-matter coupling strength~\cite{wigger2017exploring} (SI).

\subsection{Four-wave mixing spectra}\label{sec:FWM}
~

\noindent
To get a first expectation for the avoided crossing behaviour in FWM, we apply the model used to describe the PL spectra in Fig.~\ref{fig:PL} and calculate the FWM spectra for ultrafast laser pulses in the $\delta$-pulse limit depicted in Fig.~\ref{fig:FWM_sim}, where we consider vanishing pulse delays $\tau_{12}=\tau_{23}=0$ and a dephasing rate of $\gamma=0.05$~meV. Note, that the electric field is here given as energy shift between the lowest hole states in the two dots. We also calculated corresponding spectra by solving the Lindblad equation considering a non-vanishing pulse duration of $\Delta t=0.2$~ps and achieved a very similar result (SI). Here, we find the same structure of avoided crossings as marked by the arrows. The lower energy one at $F\approx-0.75$~meV is bridged by a spectral line, such that one avoided crossing remains at $F=0$~meV. In the higher energetic transitions we find three clear gaps in the spectrum (arrows), two around $F\approx-0.75$~meV and the other at $F=0$~meV. In addition, the intensity distribution of the different lines has to be addressed. Due to the thermal occupation ($T=7$~K, as in the experiment) of the two ground (hole) states before any optical excitation, the resulting intensities depend on which of the QDs has the lowest hole state energy for a given applied electric field. This finally results in brighter FWM lines at lower energies for larger bias values and in higher intensities at larger transition energies for smaller bias values. The visibility of the crossing lines around $F=-0.4$~meV [$\Delta V\approx-730$~mV in Fig.~\ref{fig:PL}(b)] is significantly reduced in FWM compared to PL (Fig.{\ref{fig:PL}(b)). This happens because of the nonlinear character of the signal in the $\chi^{(3)}$ regime, which suppresses transitions with weaker dipole matrix elements.\\
\begin{figure}[t!]
    \centering
    \includegraphics[width=\columnwidth]{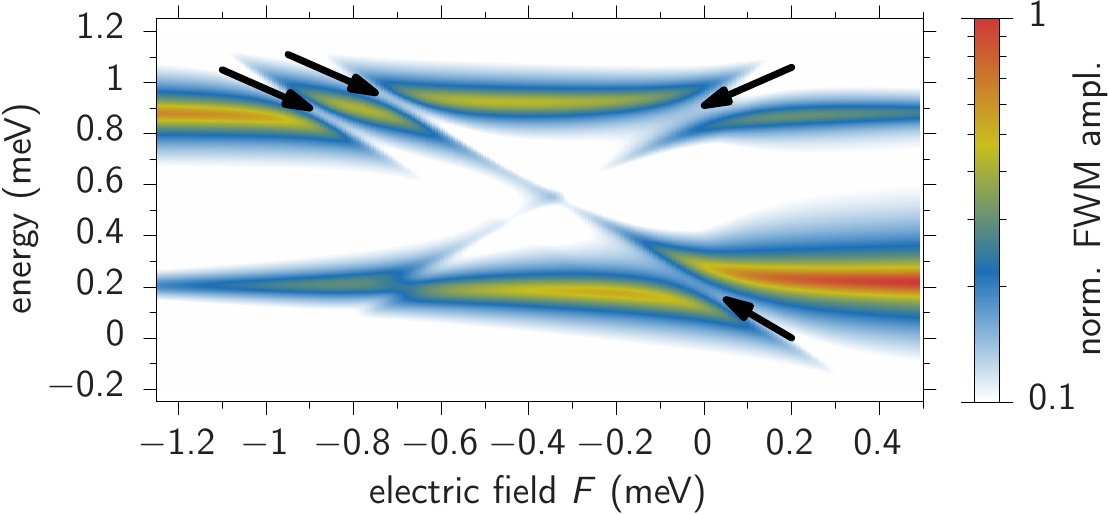}
   \caption{Simulated bias scan of the FWM spectrum of the QDM corresponding to the dashed lines in Fig.~\ref{fig:PL}(a). The arrows mark the clear avoided crossings.}
    \label{fig:FWM_sim}
\end{figure}%
In Fig.~\ref{fig:FWM_exp}(a) we show the respective measured bias scan of the FWM spectra for small pulse delays $\tau_{12}=\tau_{23}=0.1$~ps which maximize the FWM response and can still be treated as $\tau_{12}=\tau_{23}=0$ in the simulation. Note, that the depicted bias range differs from the PL measurement in Fig.~\ref{fig:PL}(a). The reason is that the state energies strongly depend on the optical excitation conditions (see SI for details). Therefore, the bias ranges in PL and FWM cannot be compared directly. The lowest energetic line stems from a neutral exciton. Therefore, it involves different charge states of the system and does not interact with the trion transition lines. We thus use it to phase-correct the delay scans, to generate 2D FWM spectra (see Fig.~\ref{fig:FWM_exp}(b, c) and SI). The upper two spectral lines clearly belong to the trion transitions we are interested in. Obviously, much fewer spectral features are visible compared to the measured PL spectra [Fig.~\ref{fig:PL}(a)], which is again related to the fact that the FWM signal is nonlinear and stems from the $\chi^{(3)}$ response of the optical transitions. Therefore, one naturally expects spectral lines that are weaker in PL to be strongly suppressed in FWM. Consequently, they might drop below the signal-to-noise level and do not appear in the detected spectra. Nevertheless, by comparing the measurement in Fig.~\ref{fig:FWM_exp}(a) with the simulation in Fig.~\ref{fig:FWM_sim} we find striking similarities. First of all, we observe the same intensity distribution with stronger low-energy lines at larger electric fields ($\Delta V\approx -200$~mV) and stronger high-energy lines at smaller electric fields ($\Delta V\approx-450$~mV). We also see one small indication of an avoided crossing in the low-energy line at $\Delta V\approx-300$~mV ($\nwarrow$ arrow), while no such feature appears in this spectral line at smaller fields. We further find a few indications of avoided crossings in the transition line at higher energies. Although only slightly above the noise-level, we can identify three gaps in the spectral line, namely at $\Delta V=-390$, $-360$ ($\searrow$ arrows), and $-300$~mV ($\swarrow$ arrow). These observations are in agreement with the predictions from the model in Fig.~\ref{fig:FWM_sim}. Our $\bm{k}\cdot \bm{p}$ model with an extended electron and hole state basis and a QDM morphology mapped -- as far as possible -- from a typical grown QDM structure yields FWM spectra that exhibit exactly the same features. It shows two horizontal branches interrupted by avoided crossings, and their intensities decay in the opposite bias directions and no additional transitions are visible (SI). This validates the applicability of the semi-empirical model, but may open the path to study the effect of composition and morphology on the nonlinear optical response and quantum coherence properties of the QDM.\\
\begin{figure}[t!]
    \centering
    \includegraphics[width=\columnwidth]{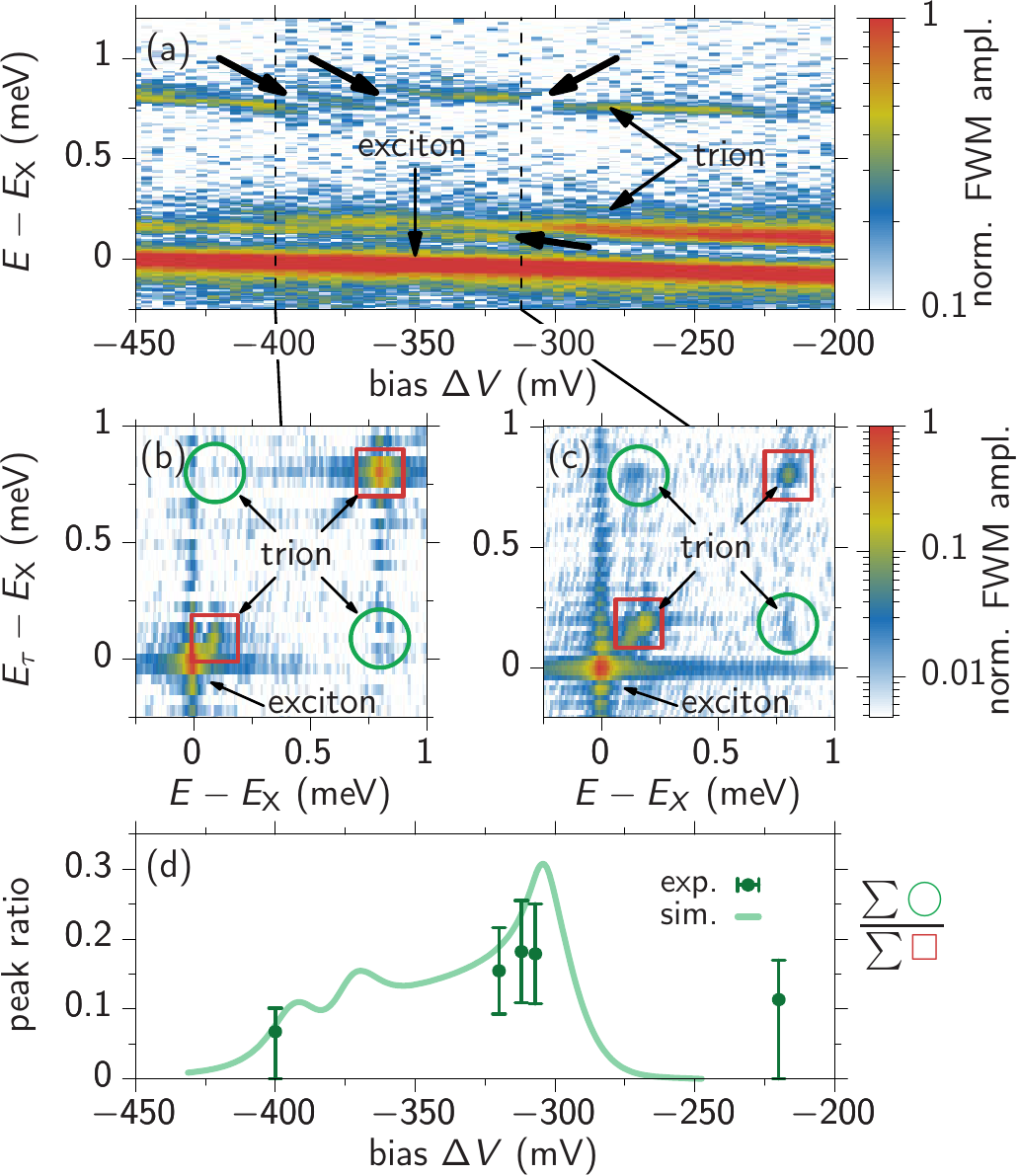}
   \caption{(a) Measured bias scan of the QDM FWM spectra with pulse delays $\tau_{12}=\tau_{23}=0.1$~ps. The arrows correspond to those in Fig.~\ref{fig:FWM_sim}(b). (b, c) Exemplary 2D FWM spectra at the bias values marked in (a). Red boxes mark the diagonal peaks and green circles the off-diagonal peaks used to determine the peak ratios in (c). (d) Peak ratios in the 2D FWM spectra to quantify the coherent coupling as a function of applied bias. Experiment as dots and simulation as line with homogeneous and inhomogeneous dephasing rates $\gamma=0.05$~meV and $\sigma=0.15$~meV, respectively.}
    \label{fig:FWM_exp}
\end{figure}%
We move forward to the main result, which is the demonstration and control of the quantum coherence transfer between different trion complexes in the QDM. The bias-controlled coherent coupling of the QDM is revealed by measuring 2D FWM spectra, as exemplarily depicted in Figs.~\ref{fig:FWM_exp}(b, c). The two energy axes of these 2D spectra are on the one hand determined by the spectrometer in the emitted signal (horizontal, $E$) and on the other hand by a Fourier transform with respect to the delay $\tau_{12}$ in the FWM measurement (vertical, $E_\tau$). The meaning of the different peaks in the maps can directly be read from the plots: Optical absorption and FWM emission from the same trion is represented by the peaks on the diagonal of the plot (red squares). Off-diagonal peaks, at the positions marked by the green circles, stem from an optical absorption from one trion ($E_\tau$ axis) and FWM generation from another one ($E$ axis), due to their coherent tunnel-coupling. Therefore, the presence of these peaks in Fig.~\ref{fig:FWM_exp}(c) confirms the coherent quantum state transfer for the electric field, where the two hole/ground states are in resonance ($\Delta V\approx-300$~mV). We see that the off-diago\-nal peaks are suppressed in Fig.~\ref{fig:FWM_exp}(b) ($\Delta V=-400$~mV), indicating a much less efficient coherence transfer for electric fields where the excited states are not in close resonance.\\
Despite the limited signal-to-noise ratio of the 2D spectra, which makes a consistent quantitative analysis challenging, we carefully determine the relative strength of the off-diagonal peaks with respect to the diagonal ones, and therefore quantify the coherent coupling between the QDs. Special attention has to be payed to disregard the impact of the dominant neutral exciton peak, marked in the plots (SI). We then determine the peak visibility by calculating the ratio of the sum-intensity of the off-diagonal peaks with respect to the diagonal ones, as symbolically written next to Fig.~\ref{fig:FWM_exp}(d). In this figure the peak ratios from the performed experiments are shown as green dots and the corresponding simulation as pale green curve. For the simulation we consider a similar spectral broadening as in Fig.~\ref{fig:FWM_sim} (SI). Both, experiment and theory show increased peak ratios around the avoided crossing at a bias of $\Delta V\approx-300$~mV and the measurement and the simulation show a reasonable agreement. Note, that to match the experiment we have to assume additional inhomogeneous dephasing due to fluctuations of the transition energies with a standard deviation of $\sigma = 0.15$~meV (see SI for details)~\cite{kasprzak2022coherent}. In short, an additional dephasing of the cross-coherences $\left|T_{2,1}\right>\!\left<h_{1,2}\right|$, exceeding the pure dephasing acting on the optical active coherences $\left|T_{1,2}\right>\!\left<h_{1,2}\right|$, will reduce the visibility of the off-diagonal peaks in the 2D spectrum. Such an additional dephasing is expected for coherences between states differing in charge distribution due to their large sensitivity to environmental fluctuations.
\begin{figure}[b!]
    \centering
    \includegraphics[width=\columnwidth]{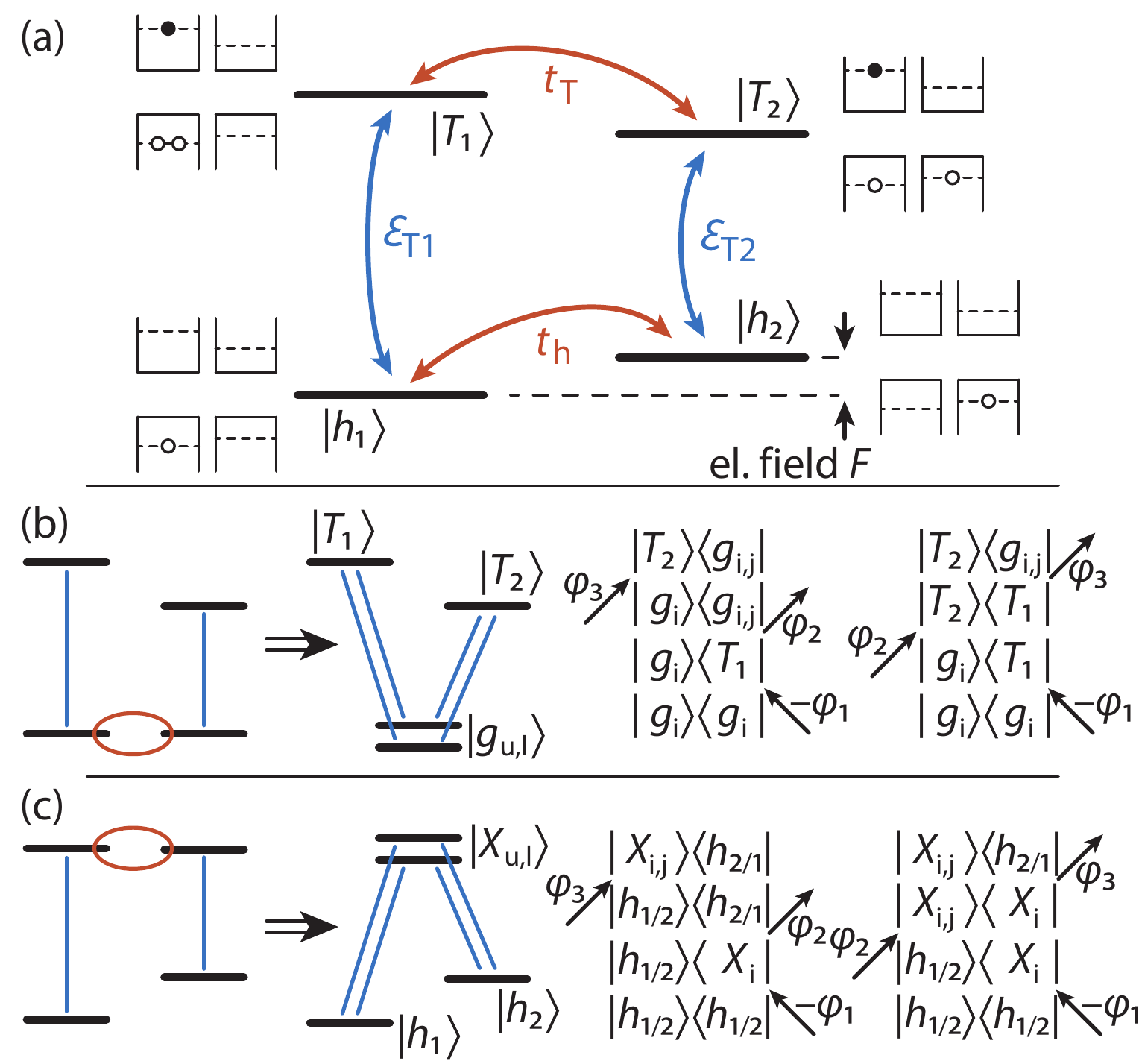}
   \caption{(a) Schematic picture of the minimum model to explain the coherent coupling effect. We include two hole $\ket{h_{1,2}}$ and two trion states $\ket{T_{1,2}}$ in the two QDs as depicted next to each level. We consider an optical transitions $\E_{\rm T1,T2}$ (blue) and tunnel couplings between the hole and the trion states (orange). (b, c)~Illustration of the hole (b) and trion resonances (c). Left: Exciton level structure with optical transitions (blue lines) and tunnel coupling (orange ellipses). Middle: New eigenstates forming ground $\left| g_\text{u,l}\right>$ (b) and excited states $\left| X_\text{u,l}\right>$ (c) doublets of upper (u) and lower (l) states. Right: Relevant double-sided Feynman diagrams leading to the off-diagonal (coherent coupling) peaks in the 2D spectra with $\text{i},\text{j}=\text{u},\text{l}$.}
    \label{fig:scheme}
\end{figure}

\subsection{Minimum model interpretation}\label{sec:mini}
~

\noindent
To fully understand the transfer of the coherence between the trions in the FWM process we developed a minimum model that only considers one ground hole state per QD ($\left|h_{1,2}\right>$) and the corresponding bright trion states with an additional electron-hole pair in one of the dots ($\left|T_{1,2}\right>$), which makes a total of four levels similar to Ref.~\cite{langer2012magnetic,salewski2017high}. A schematic of the coupling mechanisms between these states is depicted in Fig.~\ref{fig:scheme}(a). While optical excitations $\E_{\rm T1,T2}=\E(t)M_{\rm T1,T2}$ are only possible within each trion complex (T1, T2, vertical connection), the tunnelings $t_{\rm h,T}$ connect the hole (h) and trion (T) states (horizontal connection), respectively. This system can be described by the Hamiltonian:
\begin{align*}
    H=&\ pF \proj{h_2}{h_2} + E_1 \proj{T_1}{T_1} + (pF+E_2) \proj{T_2}{T_2}  \\
    &+ \E_{\rm T1}^*(t) \proj{h_1}{T_1} + \E_{\rm T1}(t) \proj{T_1}{h_1} \\
    &+ \E_{\rm T2}^*(t) \proj{h_2}{T_2} + \E_{\rm T2}(t) \proj{T_2}{h_2}\\
    &+ t_{\rm h} (\proj{h_1}{h_2} + \proj{h_2}{h_1}) + t_{\rm T} (\proj{T_1}{T_2} + \proj{T_2}{T_1} )\,.
\end{align*}%
When we now assume that the two hole states are in resonance we can neglect the tunneling between the trion states, because they are far from resonance ($t_{\rm T}=0$). This situation is schematically shown in Fig.~\ref{fig:scheme}(b, left) where the tunnel coupling between the hole states leads to a hybridisation resulting in a new upper $\left|g_\text{u}\right>$ und lower ground state $\left|g_\text{l}\right>$ (middle). The optical active transitions now connect both of these states to each trion state resulting in a V-level structure. To construct the FWM signals with the phase combination $\varphi_3+\varphi_2-\varphi_1$ we consider the two double-sided Feynman diagrams on the right. We see that the V-level structure induced by the hole tunnelling results in different possibilities to have an absorption in $\left| T_1\right>$ and an emission in $\left| T_2\right>$ (the other off-diagonal peak is retrieved by exchanging 1 and 2). The 1st diagram reaches the coherence transfer between the two trion complexes via the ground state doublet $\proj{g_\text{i}}{g_\text{i,j}}$ after $\varphi_2$ and the 2nd one via the inter-trion coherence $\proj{T_2}{T_1}$. It is interesting to note, that coherent coupling is mediated via the common ground states where the hole states are de-localised across the two dots, but in the trion states the hole states are localised in each dot. The lifted degeneracy of the ground states in theory leads to the generation of multiple peaks on the off-diagonal, which cannot be resolved in the experiment but are visible in the simulated 2D spectra in the SI. An equivalent situation is found when the trion states are in resonance forming new excited states $\left|X_\text{u,l}\right>$ resulting in an $\Lambda$-level scheme as depicted in Fig.~\ref{fig:scheme}(c). Here, the initial state is a thermal mixture of the hole states $\left|h_1\right>$ and $\left|h_2\right>$ and the Feynman diagrams show that the path of coherence transfer goes via the hole states $\proj{h_1}{h_2}$ or the excited states doublet $\proj{X_\text{u}}{X_\text{l}}$ to the respective other hole state.\\
Before the FWM pulse sequence is started, the QDM is typically in its ground states, i.e., in a thermal equilibrium of the hole states. However, recently an experimental scheme has been developed to deterministically initialise the QDM in one of the two hole states~\cite{bopp2022quantum}. It is thus becoming possible to induce a uni-directional coherence transfer between the two QDs. This would manifest in the 2D FWM spectrum with the appearance of only one off-diagonal peak. One would be starting from a non-eigenstate of the system which introduces additional dynamics in the system. It is important to note, that this FWM scheme offers the possibility to probe different coherence dynamics that are not optically active, namely $\proj{g_\text{u}}{g_\text{l}}$, $\proj{h_1}{h_2}$, $\proj{X_\text{u}}{X_\text{l}}$, and $\proj{T_1}{T_2}$. To achieve this one would have to vary the delay between the 2nd and 3rd pulse ($\tau_{23}$) as can be seen from the Feynman diagrams in Fig.~\ref{fig:scheme}(b, c).

\section{Conclusions}
In summary, revealing controlled coherent coupling between different trion complexes in a QDM renders a major advance in the field of photonic quantum technology. Our proof-of-principle demonstration of nonlinear multi-wave mixing spectroscopy on these systems opens the door to address coherences that are not optically active, making them significantly longer lived than their bright counter parts, and could therefore be used as quantum storage~\cite{langer2012magnetic, salewski2017high}. To improve the performance of quantum devices, current work aims to generate singlet-triplet qubits~\cite{kim2011ultrafast, greilich2011optical, weiss2012coherent}, which are expected to be robust against decoherence from electrostatic and magnetic fluctuations in the solid. Due to its selective access to different quantum coherences, the FWM method stands out as an excellent tool to exploit this new benchmark for quantum computing with solid state qubits.


\section*{Acknowledgements}  
D.W. thanks the Polish National Agency for Academic Exchange (NAWA) for financial support within the ULAM program (Grant No. PPN/ULM/2019/1/00064) and the Science Foundation Ireland (SFI, Grant No. 18/RP/6236). J.K. acknowledges the support from MCQST and Global Invited Professorship at the TU Munich and 'Tandem for Excellence' IDUB scheme at the University of Warsaw. P.Mr., M.K, K.G., C.D., F.B., J.J.F, and P.Ma. acknowledge support from the Polish National Science Centre (NCN) (Grant No. 2016/23/G/ST3/04324). The Berlin team acknowledge support by the German Federal Ministry of Education and Research (BMBF) through the pro\-jects Q.Link.X and QR.X. We thank Thilo Hahn for helpful comments regarding inhomogeneous broadening.


%
\section*{References}

\end{document}


\maketitle

\begin{affiliations}
\item {\small Institute of Theoretical Physics, Wroc\l{}aw University of Science and Technology, 50-370~Wroc\l{}aw, Poland}
\item {\small School of Physics, Trinity College Dublin, Dublin 2, D02 PN40, Ireland}
\item {\small Institute of Solid State Physics, Technische Universit\"at Berlin, 10623 Berlin, Germany}
\item {\small Lehrstuhl f\"ur Angewandte Festk\"orperphysik Ruhr-Universit\"at Bochum, 44780 Bochum, Germany}
\item {\small Laboratory for Optical Spectroscopy of Nanostructures, Department of Experimental Physics, Wroc\l{}aw University of Technology, 50-370 Wroc\l{}aw, Poland}
\item {\small Walter Schottky Institut and Physik Department, Technische Universit\"at M\"unchen, 85748 Garching, Germany}
\item {\small Department Physik, Universit\"at Paderborn, 33098 Paderborn, Germany}
\item {\small Universit\'e Grenoble Alpes, CNRS, Grenoble INP, Institut N\'{e}el, 38000 Grenoble, France}
\item[$^\ast$] {\small daniel.wigger@tcd.ie, jacek.kasprzak@neel.cnrs.fr}
\end{affiliations}

\begin{abstract}\small
This document contains:
\noindent \begin{itemize}
\item[\textbf{S1.}] \textbf{Four-wave mixing experiment}\\[-10mm]
	\begin{itemize}
		\item[\textbf{S1A.}] \textbf{Experimental method and device}\\[-9mm]
		\item[\textbf{S1B.}] \textbf{Coherence dynamics}\\[-9mm]
		\item[\textbf{S1C.}] \textbf{Rabi rotations}\\[-9mm]
		\item[\textbf{S1D.}] \textbf{Identification of the neutral exciton-biexciton complex}\\[-9mm]
		\item[\textbf{S1E.}] \textbf{Power dependence of PL spectra}\\[-9mm]
		\item[\textbf{S1F.}] \textbf{Fitting 2D FWM spectra}
	\end{itemize}
	
\item[\textbf{S2.}] \textbf{Theory}\\[-10mm]
	\begin{itemize}
		\item[\textbf{S2A.}] \textbf{Semi-empirical model}\\[-9mm]
		\item[\textbf{S2B.}] \textbf{Calculation of 2D spectra for ultra-short pulses}\\[-9mm]
		\item[\textbf{S2C.}] \textbf{Dynamical simulations}\\[-9mm]
		\item[\textbf{S2D.}] \textbf{Inhomogeneous dephasing}\\[-9mm]
		\item[\textbf{S2E.}] \textbf{\kp model}
	\end{itemize}
\end{itemize}
\end{abstract}

\newpage
\footnotesize
\section{Four-wave mixing experiment}
In this section we present details on the performed experiment and used sample (Sec.~S1A). We show standard characterization measurements of the coherence (Sec.~S1B) and light-matter coupling (Sec.~S1C). Further, we identify the neutral exciton transition (Sec.~S1D), demonstrate the power dependence of the photoluminescence (PL) spectra (Sec.~S2E), and describe our fitting procedure for the 2D spectra (Sec.~S1F).

\subsection{S1A. Experimental method and device}\label{sec:exp}
~

\begin{figure}[h]
    \centering
    \includegraphics[width=0.45\columnwidth]{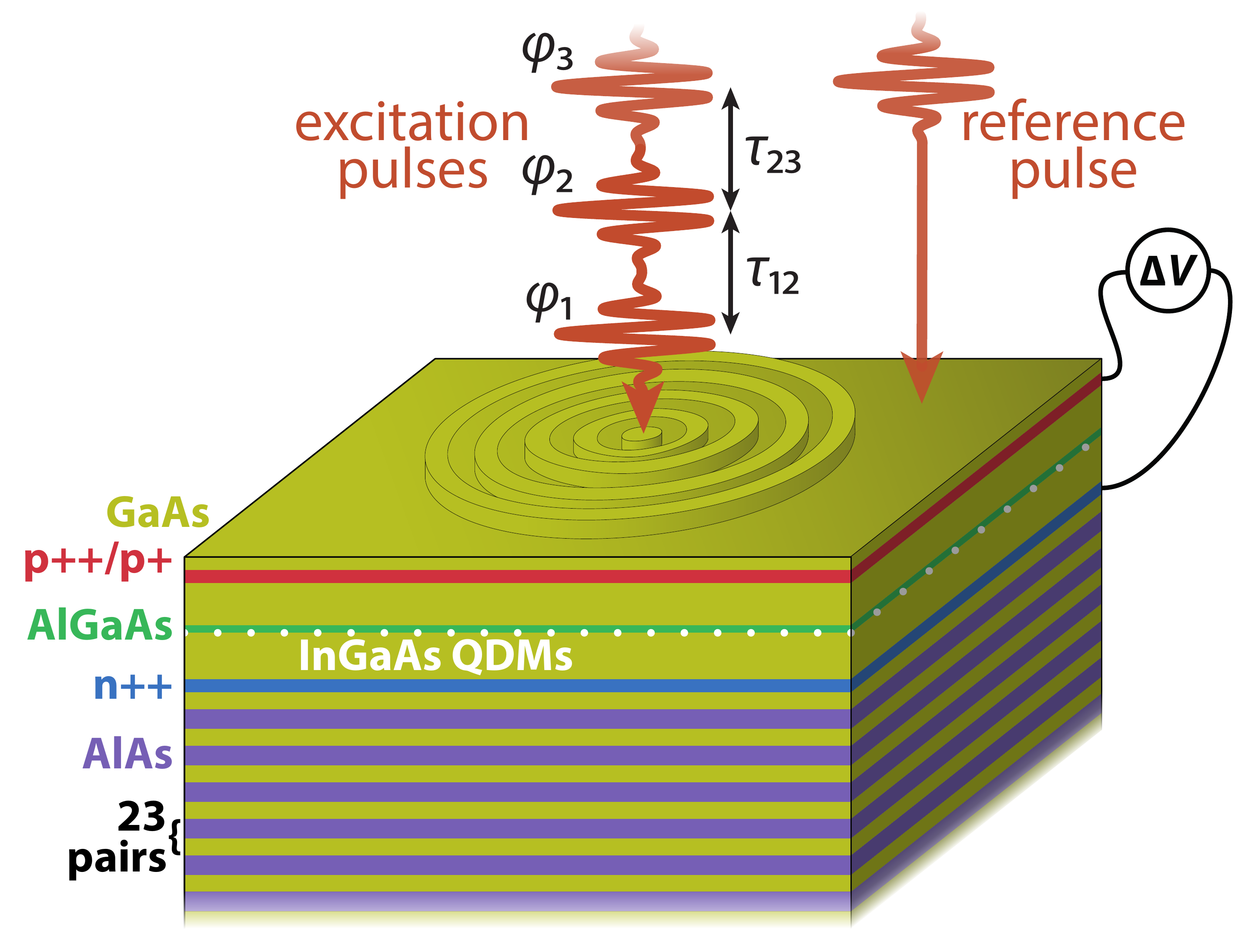}
   \caption{Schematic picture of the sample structure. Above a GaAs/AlAs distributed Bragg structure the InGaAs quantum dot molecule (QDM) layer is sandwiched between charge-doped layers, which are connected to a bias source. For details on the sample design and processing we refer to Ref.~\cite{SchallAQT21}~\!. The optical in- and out-coupling is improved by a circular Bragg grating, also called a bulls-eye photonic structure, on the top surface. The laser pulses used for the optical excitation in the four-wave mixing experiment are focused on the center of the grating, while the reference beam hits the free surface in the vicinity.}
    \label{fig:sample}
\end{figure}

\noindent
We employ the heterodyne spectral interferometry technique introduced in
Ref.~\cite{LangbeinOL06} to detect four-wave mixing (FWM) signals from the quantum dot molecule (QDM). Its two-beam version was
described in detail in the Supplementary Material of Ref.~\cite{KasprzakNPho11}~\!, while its
extension to a three-beams configuration was presented in the Supplementary Material of
Ref.~\cite{FrasNatPhot16}~\!. We use a Ti:Sapphire femto-second oscillator tuned into
resonance with bright excitonic (trion) transitions of the QDM around 930~nm (1333~meV). In order
to avoid driving spectrally nearby transitions, for instance biexcitons, the pulses are
spectrally shaped (with a passive, diffraction grating-based pulse shaper) increasing
their duration to $\approx 0.4$~ps (full width at half maximum of the intensity spectrum). The main beam is then split into the reference pulse $\E_{\rm R}$
and three excitation pulses $\E_{1,2,3}$. In the latter, the consecutive pulses in the
train are distinctly phase-shifted using acousto-optic modulators driven at the
radio-frequencies ($\Omega_1, \Omega_2, \Omega_3)=(80, 79, 79.77)$~MHz. Using mechanical
delay stages, they then acquire respective delays $\tau_{12}$  and $\tau_{23}$, between
the first two and last two arriving beams, respectively. $\E_{1,2,3}$ are recombined into
the same spatial mode and propagate unidirectionally. Using an external microscope
objective, $\E_{1,2,3}$ are focused on the sample surface, impinging the central part of
the circular Bragg grating, while $\E_{\rm R}$ is placed on the flat surface around
15~\textmu m apart (Fig.~\ref{fig:sample}). The signal is collected in the reflection
direction. The reflection from the QDM, which also contains the FWM signal,
is interfered with $\E_{\rm R}$ and heterodyned at the frequency
$\Omega_3+\Omega_1-\Omega_1$ in the same manner as explained in previous
publications~\cite{LangbeinOL06, KasprzakNPho11, FrasNatPhot16, HahnAdvSci21}. Note, that in this work we apply the degenerate two-pulse FWM scheme, i.e., $\tau_{23}=0$.\\
The sample is kept in a He-flow cold-finger cryostat at the temperature of 7~K. To tune the spectral
separation between quantum levels in the QDM, we vary the gate voltage $\Delta V$
(Fig.~\ref{fig:sample}), supplied by an external voltage source micro-bonded to the
electrodes of the QDM device via the feed-through connectors of the cryostat. In the
initial part of the experiment, the current-voltage characterization was carried out (not shown). To
assure a safe operation of the p-i-n diode the current is limited to $80$~\textmu A. The
voltage source, provided by {\it Keithley}, is interfaced via the General Purpose Interface Bus (GPIB) with the
home-developed software controlling the whole experiment.

\subsection{S1B. Coherence dynamics}\label{sec:coh}
~

\begin{figure}[h]
    \centering
    \includegraphics[width=0.5\columnwidth]{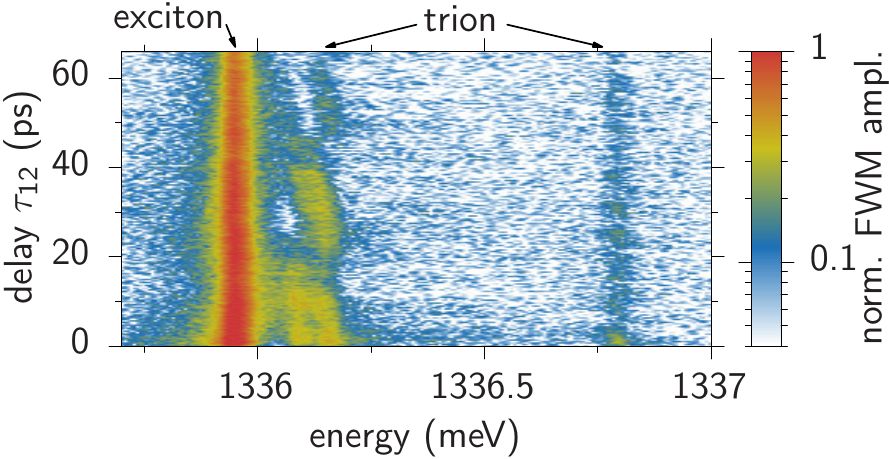}
   \caption{Spectral dynamics of the trion transitions as a function of the delay $\tau_{12}$. The trion lines show a coherence beating due to the coherent coupling. The exciton line just decays due to the decoherence of the zero phonon line.}
    \label{fig:spec_dyn}
\end{figure}

\begin{figure}[h]
    \centering
    \includegraphics[width=0.5\columnwidth]{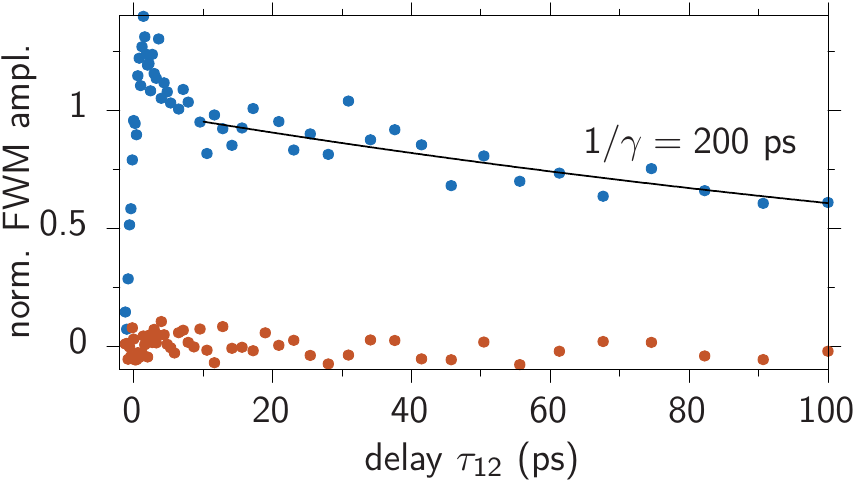}
   \caption{Spectrally integrated coherence dynamics as a function of $\tau_{12}$ of the neutral exciton line, measured far from the hole tunneling resonance (blue points). The orange points represent the noise level. The long-time decay is fitted with a single exponential with the dephasing time $200$~ps. The initial faster decay is due to phonon-induced dephasing.}
    \label{fig:deph}
\end{figure}

\noindent
Figure~\ref{fig:spec_dyn} shows an exemplary coherence dynamics measurement by varying the
delay $\tau_{12}$ in the FWM experiment. While the neutral exciton line only decays due to
dephasing processes, the trion transitions exhibit pronounced oscillations. These involved
dynamics originate from the coherent coupling between the different states and lead to the
off-diagonal peaks in the 2D FWM spectra (see e.g. Fig.~\ref{fig:2D}). \\
The beat dynamics in the FWM signal is suppressed when only one of the transitions is optically driven. For such a measurement the spectrally integrated FWM amplitude of the neutral exciton is shown in Fig.~\ref{fig:deph}. By fitting the long-time decay with a single exponential function, we deduce a dephasing time of 200~ps. This dephasing rate is used for the simulations in Fig.~\ref{fig:Lindblad}. The initial rapid drop of the FWM amplitude is most likely stemming from a phonon-induced dephasing process~\cite{wigger2020acoustic} and is not further investigated here.

\newpage
\subsection{S1C. Rabi rotations}\label{sec:Rabi}
~

\begin{figure}[h]
    \centering
    \includegraphics[width=0.5\columnwidth]{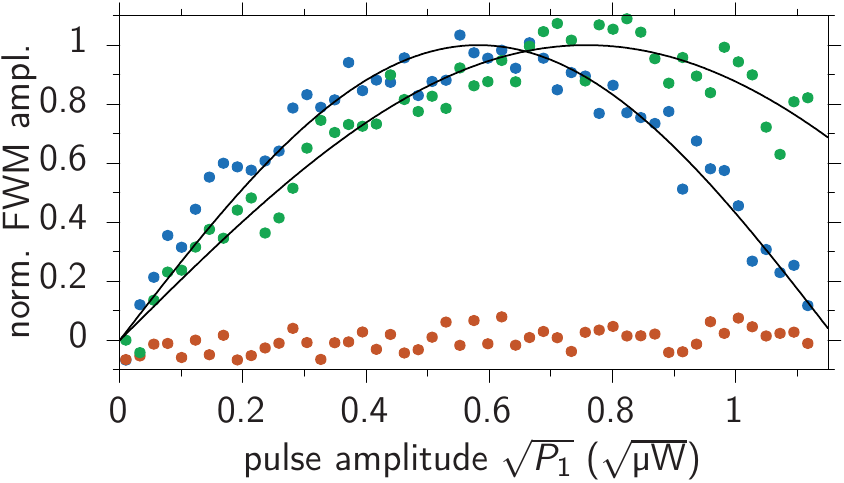}
   \caption{Rabi rotation experiment, performed on the neutral exciton (blue points) and the trion (green points) transitions. The orange points represent the noise level. Pulse areas of $\pi/2$ are found for pulse amplitudes around $0.7\ {\sqrt{\text{\textmu W}}}$.}
    \label{fig:Rabi}
\end{figure}

\noindent
To characterize the strength of the light-matter coupling of the QDM in Fig.~\ref{fig:Rabi} we plot the spectrally integrated FWM amplitude (for $\tau_{12}=0$) as a function of the applied laser pulse amplitude. We find the expected Rabi rotation behavior~\cite{wigger2017exploring}, which shows that a pulse area of $\pi/2$ is found for a pulse amplitude around $\sqrt{P_1}=0.7\, \sqrt{\text{\textmu W}}$. For all measurement in this work we restrict ourselves to small excitation powers in the linear regime of the Rabi rotation, i.e., $\sqrt{P_1}<0.05\ \sqrt{\text{\textmu W}}$.

\subsection{S1D. Identification of the neutral exciton-biexciton complex}\label{sec:BX}
~

\begin{figure}[h!]
    \centering
    \includegraphics[width=0.5\columnwidth]{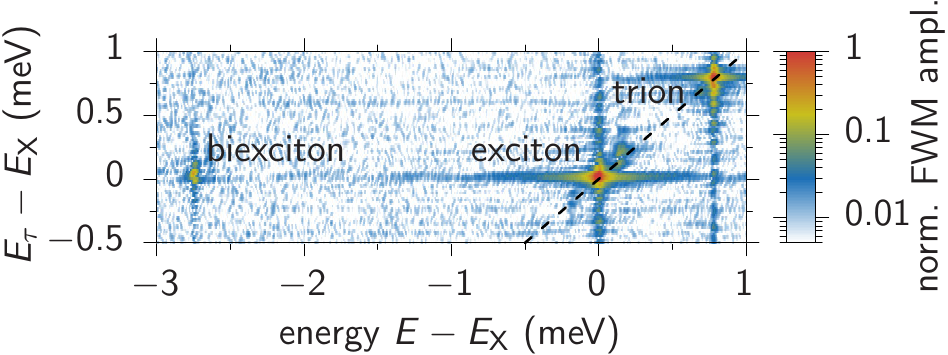}
   \caption{2D FWM spectrum showing the exciton-biexciton transition as off-diagonal peak. The exciton and trion transitions appear on the diagonal.}
    \label{fig:BX}
\end{figure}

\noindent
As stated in the main text, the bright spectral line in Fig.~3(a) stems from a neutral exciton transition. The standard proof of this type of transition is the detection of a coupled biexciton state~\cite{KasprzakNPho11}. We do this through the 2D FWM spectrum in Fig.~\ref{fig:BX}, where the biexciton appears as off-diagonal peak with a binding energy of around 2.8~meV, which is a typical value for InGaAs QDs~\cite{MermillodPRL16}.

\subsection{S1E. Power dependence of PL spectra}\label{sec:power}
~

\begin{figure}[h]
    \centering
    \includegraphics[width=0.5\columnwidth]{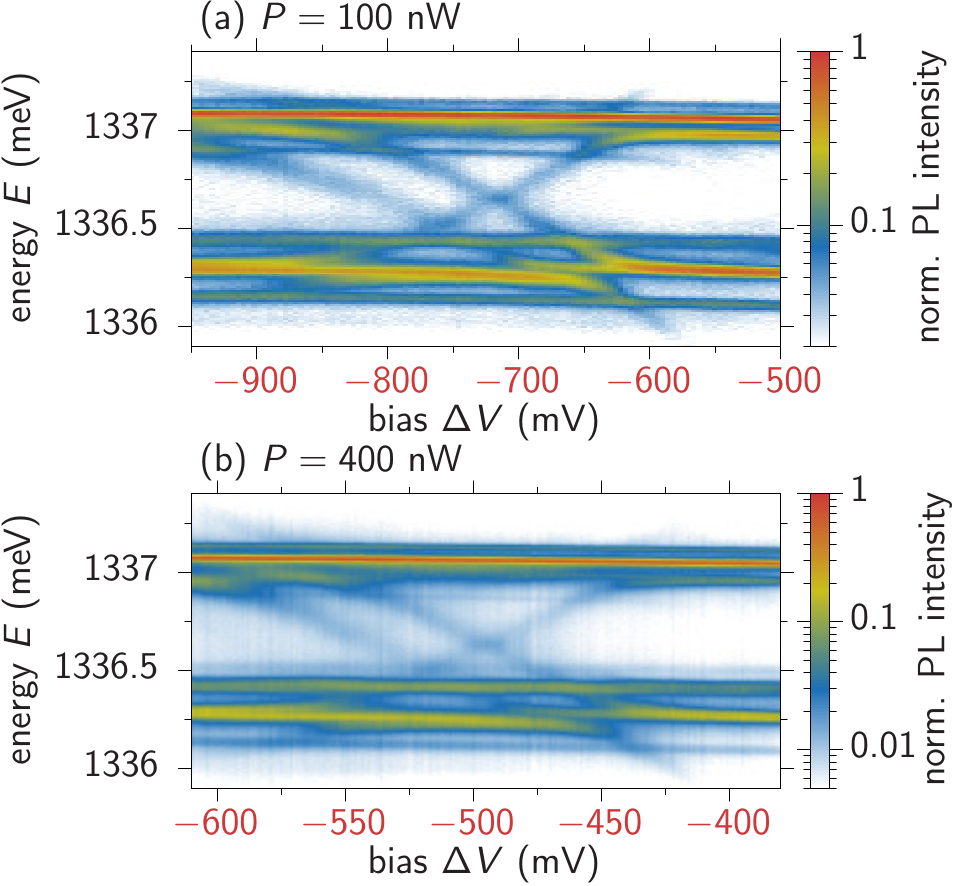}
   \caption{Bias scans of the PL intensity for two different excitation powers. (a) Same as Fig.~1(a) in the main text with $P=100$~nW. (b) 400~nW depicted for a different bias range than (a).}
    \label{fig:PL_power}
\end{figure}

\noindent
We have seen in the main text, that the bias ranges where the hole and trion resonances appear differ between the presented PL and FWM measurements. In Fig.~\ref{fig:PL_power}(a) we show the same data as in Fig.~1(a) in the main text for an optical excitation power of $P=100$~nW and in (b) the same measurement for $P=400$~nW. We find the same spectral features of line-shifts and avoided-crossings but on different bias intervals. The range is not only shifted but also scaled by almost a factor of~2. This shows that the bias-dependence of the different state energies strongly depends on the specific optical excitation conditions. We also note that even more different scenarios are observed upon additional white light illumination (not shown). Varying optical excitation, changes the free carrier density and thus also the energetic level structure in the p-i-n diode and also the alignment of the quantized levels in the QDM itself. Therefore, because already the general optical excitation schemes in PL and FWM are off-resonant and resonant, respectively, and consequently entirely different, the two experiments cannot be quantitatively compared regarding their bias range properly.

\subsection{S1F. Fitting 2D FWM spectra}\label{sec:fit}
~

\begin{figure}[t!]
    \centering
    \includegraphics[width=0.5\columnwidth]{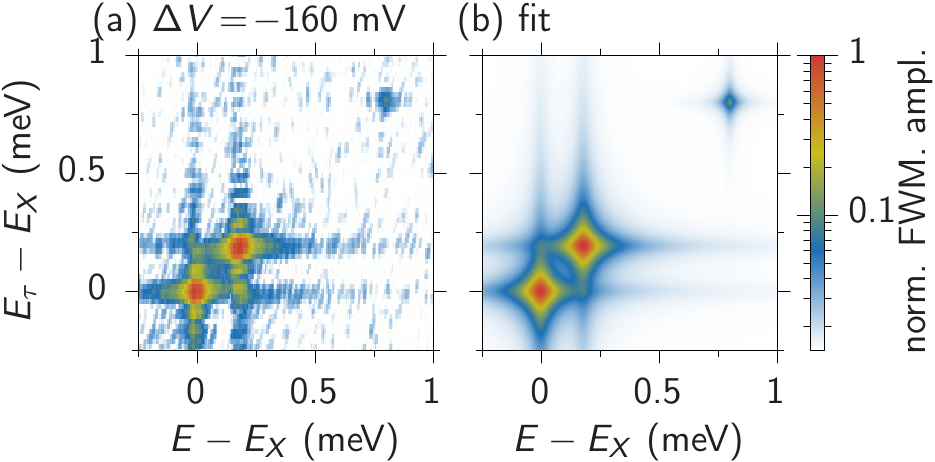}\hfill
    \includegraphics[width=0.5\columnwidth]{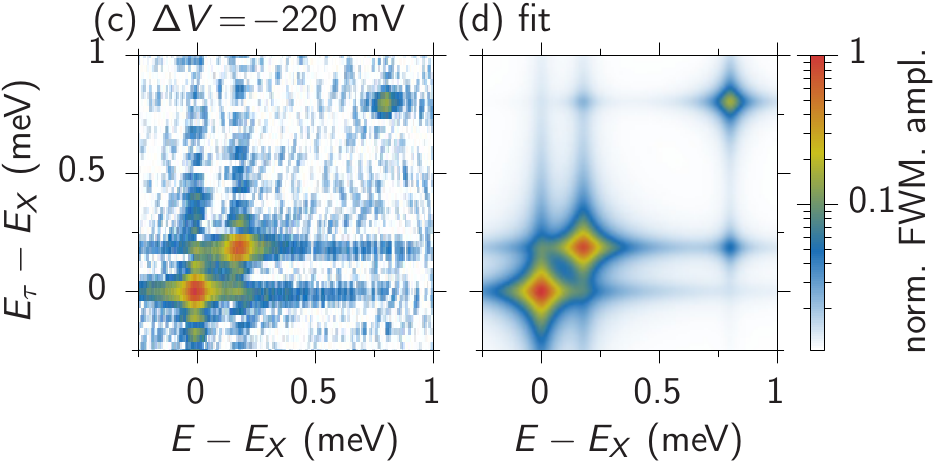}\\
    \includegraphics[width=0.5\columnwidth]{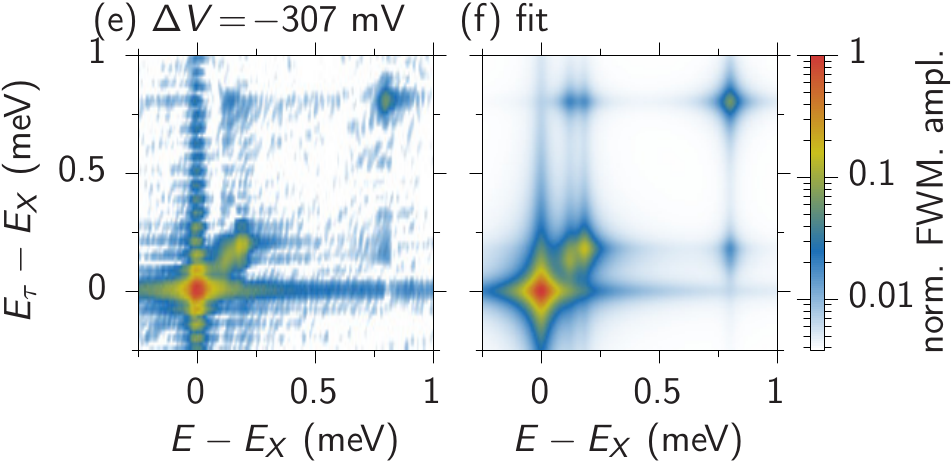}\hfill
    \includegraphics[width=0.5\columnwidth]{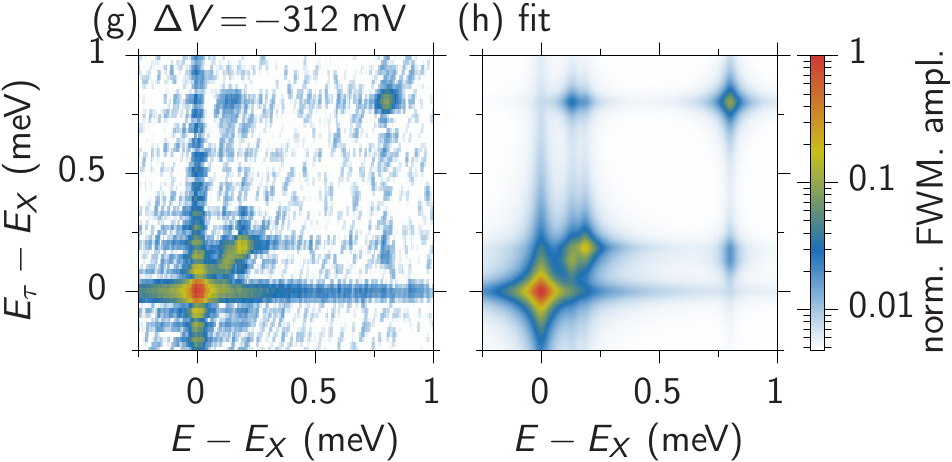}\\
    \includegraphics[width=0.5\columnwidth]{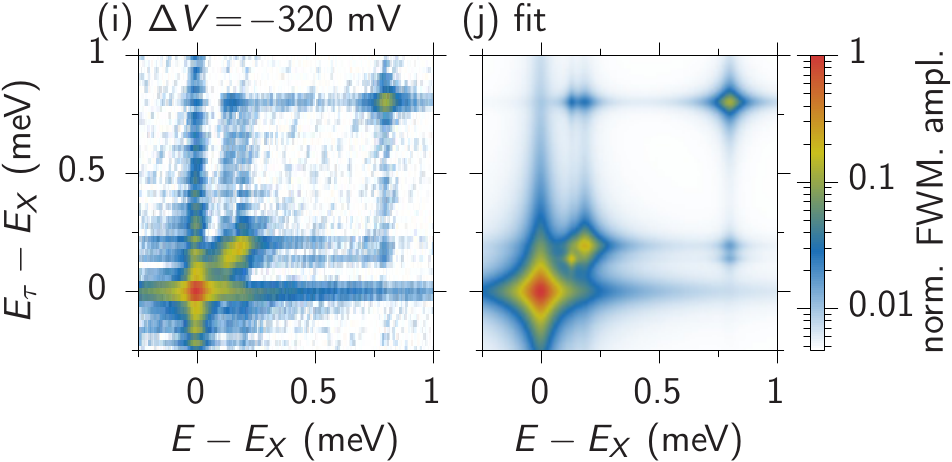}\hfill
    \includegraphics[width=0.5\columnwidth]{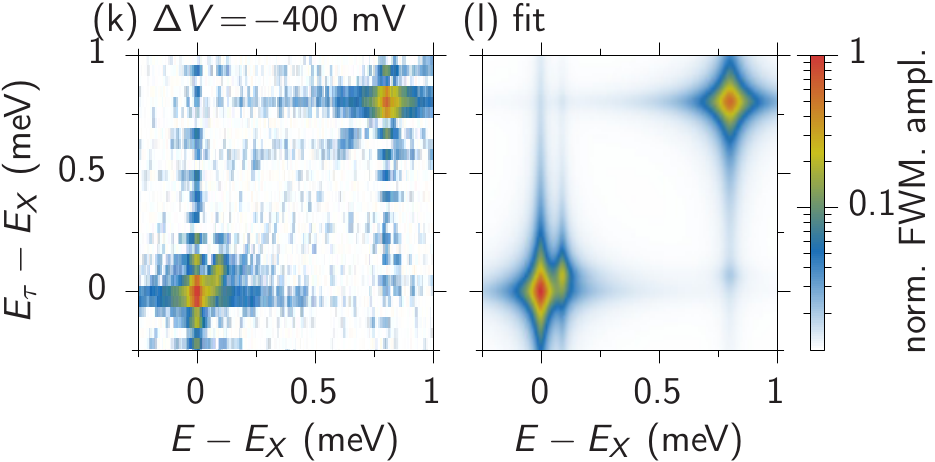}\\
 \flushleft{\includegraphics[width=0.5\columnwidth]{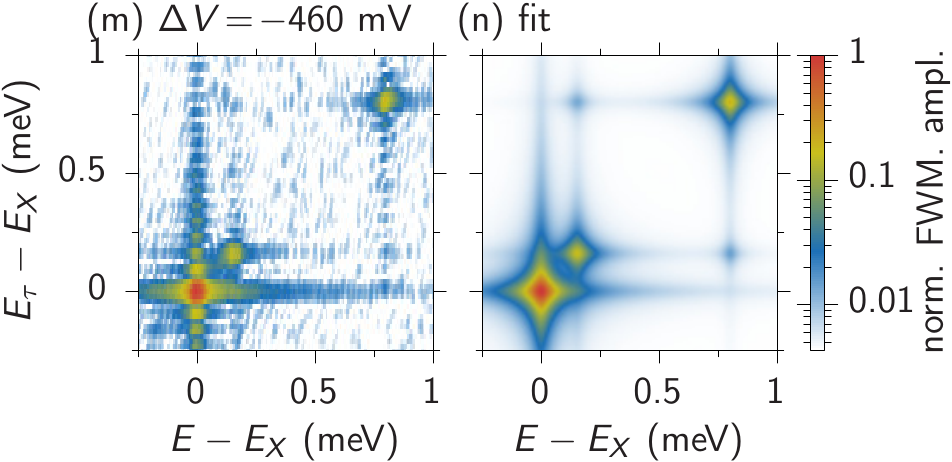} 
 \hspace{1.2cm} \includegraphics[width=0.32\columnwidth]{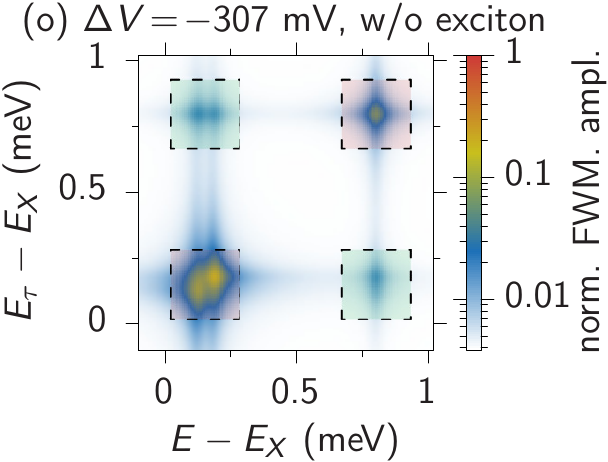}}
   \caption{(a)-(n) Measured 2D FWM spectra with respective fits for different bias values. The fit function is a sum of 2D Lorentzians. (o) Fitted 2D FWM spectrum for $\Delta V=-307$ mV without the neutral exciton line. The red shaded areas mark the integration range for the diagonal contributions, the green areas those for the off-diagonal ones.}
    \label{fig:2D}
\end{figure}

\noindent
To efficiently remove the prominent impact of the neutral exciton transition from the 2D spectra depicted in Fig.~\ref{fig:2D} we fit each spectrum with a sum of 2D Lorentzians. While for most studied bias values we use one diagonal peak for the exciton and two diagonal and two off-diagonal ones for the trion transitions, for the bias values $\Delta V=-307$, $-312$, and $-320$~mV we have to consider one additional diagonal and two additional off-diagonal peaks. This appearance of additional peaks in the bias range directly reflects the development of an avoided-crossing in the linear spectra (see Fig.~3(a) in the main text).\\
Our goal is to determine the peak ratio in the 2D spectra consistently in the measured and
the simulated data. Therefore, we simply focus on four parts marked by colored areas (dashed squares) in Fig.~\ref{fig:2D}(o). The signal amplitude summed over the red shaded areas gives the diagonal contributions and the sum over the green shaded areas the off-diagonal ones. This procedure can be easily automated to retrieve the peak ratios (green areas/red areas) for many bias values in the simulation, i.e., many 2D spectra. To determine the experimental peak ratios we use the fitted 2D spectra to reduce the uncertainties stemming from different discretization or spectral oscillations due to insufficiently long delay ranges.
\newpage
\section{Theory}
Here, we present the details of the semi-empirical theory for the coupled hole and trion system
(Sec.~S2A). This model yields
transition energies and amplitudes for dipole-allowed transitions that are used for
determining the FWM spectrum from the analytical formulas in the ultra-short pulse limit (Sec.~S2B) or from dynamical simulations (Sec.~S2C). Further, we discuss the influence of an inhomogeneous dephasing on the 2D spectra (Sec.~S2D). Finally, we present the \kp model and the spectra calculated from the
system's eigenstates and eigenenergies obtained in this way  (Sec.~S2E).

\newpage
\subsection{S2A. Semi-empirical model}\label{sec:semi-empirical}
~

\noindent
In the following electron and hole destruction (creation) operators are given by $a^{(\dag)}_\sigma$ and
$h^{(\dag)}_{n,\sigma}$ for the spin orientation $\sigma \in \{\u,\d\}$, respectively, and $n\in\{1,2\}$
indicates in which QD the hole is located. \\
\begin{figure}[h]
    \centering
    \includegraphics[width=0.5\columnwidth]{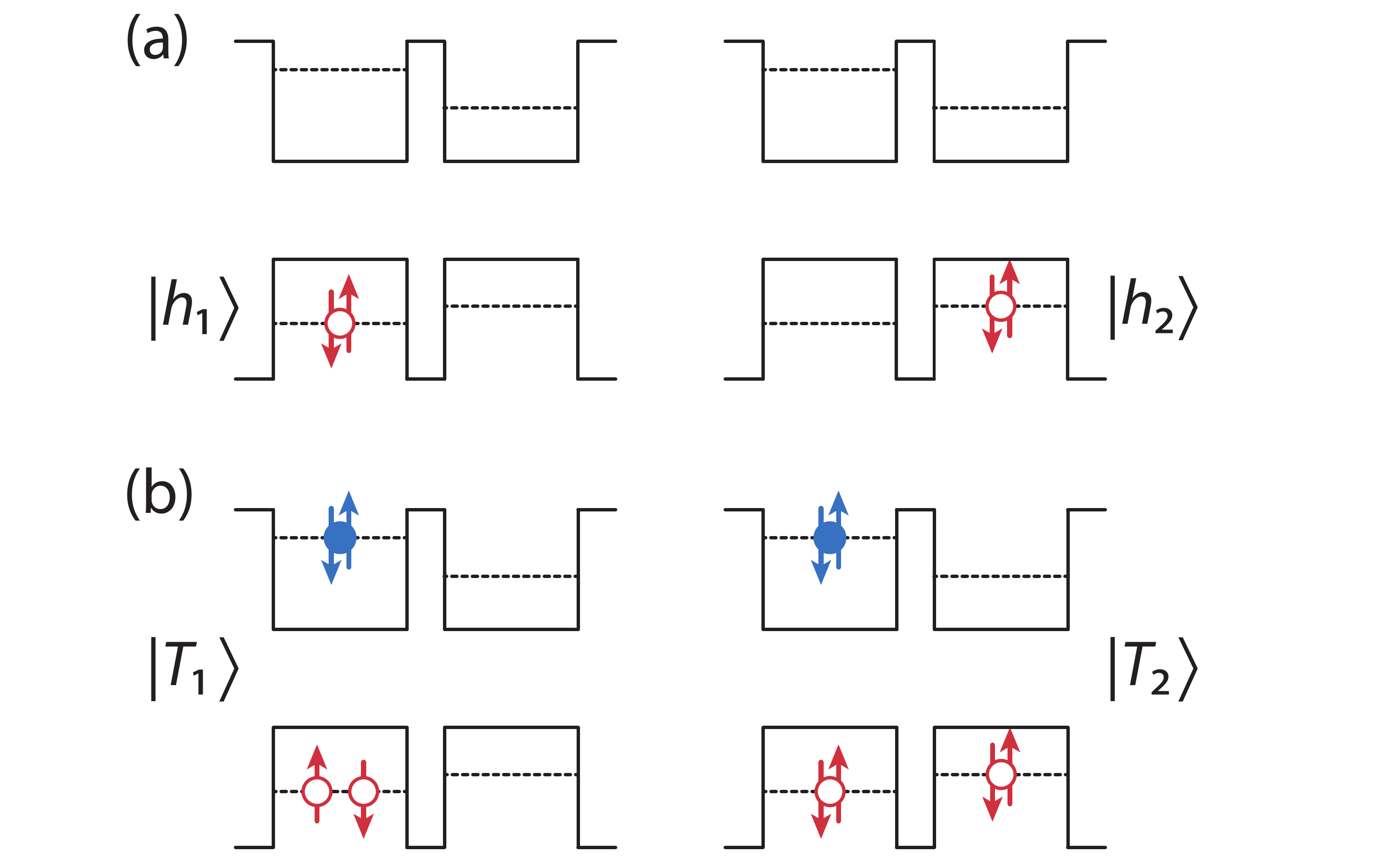}
   \caption{Schematic picture illustrating the different (a) hole and (b) trion states in the QDM taking into account the spins of the particles.}
    \label{fig:scheme_levels}
\end{figure}%
The relevant single particle states of the QDM are given by the single hole states forming
the ground states, which are characterized by the QD number and the spin orientation: 
$\ket{1_\u}, \ket{1_\d}, \ket{2_\u}, \ket{2_\d}$.
The relevant trions forming the excited states emerge by the creation of an additional
electron-hole pair, with the electron always in QD1. These three-particle complexes now
possess two holes, whose 
spin coupling results in new eigenstates that can be sorted into singlet $\ket{S_{(m,n)}}$
($m,n\in\{1,2\}$) and triplet states $\ket{T_\alpha}$ ($\alpha\in\{+,-,0\}$) defined by 
\begin{subequations}\begin{align}
	\ket{S_{(2,0)}} &= h^{\dag}_{1,\u}h^{\dag}_{1,\d}\ket{0} \\
	\ket{S_{(1,1)}} &= \frac{1}{\sqrt{2}}\left( h^{\dag}_{1,\u}h^{\dag}_{2,\d} -
 h^{\dag}_{1,\d}h^{\dag}_{2,\u} \right)\ket{0}\\
	\ket{S_{(0,2)}} &= h^{\dag}_{2,\u}h^{\dag}_{2,\d}\ket{0} \\
	\ket{T_+} &= h^{\dag}_{1,\u}h^{\dag}_{2,\u}\ket{0} \\
	\ket{T_-} &= h^{\dag}_{1,\d}h^{\dag}_{2,\d}\ket{0}  \\
	\ket{T_0} &= \frac{1}{\sqrt{2}}\left( h^{\dag}_{1,\u}h^{\dag}_{2,\d} 
+ h^{\dag}_{1,\d}h^{\dag}_{2,\u} \right)\ket{0}
\end{align}\end{subequations}
To get the full trion wave function we have to add the spin orientation $\sigma\in\{\u, \d\}$ of the remaining electron to the states such that we get $\ket{\sigma, S_{(m,n)}}$ and $\ket{\sigma, T_\alpha}$. Overall we have 4 ground/hole states and 12 excited/trion states.\\
For the minimal model introduced in the main text we may identify
$\ket{h_1}\to \ket{1_\u}$, $\ket{h_2}\to  \ket{2_\u}$, $\ket{T_1}\to \ket{S_{2,0}}$, and
$\ket{T_2}\to \ket{S_{1,1}}$ (see Fig.~\ref{fig:scheme_levels}).\\
The ground/hole state Hamiltonian reads:
\begin{align}
	H_{\rm h,\sigma} &= \sum_{\sigma=\u,\d}\left[ \left(V_2-F\right)\proj{2_\sigma}{2_\sigma} + t\left( \proj{1_\sigma}{2_\sigma} + {\rm h.c.} \right)\right],
\end{align}
where $V_{2}$ is the energy difference between the ground states in the two QDs, $F =
ed\mathcal{E}_{z}$ is the dipole energy associated with the charge displacement in the axial
electric field $\mathcal{E}_{z}$, where $d$ is the effective distance between the QDs and $e$ is the
elementary charge, and $t$ is the tunnel coupling. \\
The excited/trion state Hamiltonian reads (dropping the conserved electron spin index $\sigma$ on
the right-hand side):
\begin{align}
	H_{\rm T,\sigma} &= J \bigg( \proj{S_{(1,1)}}{S_{(1,1)}} - \sum_{+,-,0} \proj{T_\alpha}{T_\alpha} \bigg) \notag \\
	&\quad +2V_{\rm c}\proj{S_{(0,2)}}{S_{(0,2)}}  \notag \\
	&\quad + \sum_{m,n=1,2} n \left(V'_2-F\right) \proj{S_{(m,n)}}{S_{(m,n)}}  \notag \\
	&\quad + \sum_{\alpha=+,-,0} \left(V'_2-F\right) \proj{T_\alpha}{T_\alpha} \notag \\
	&\quad + t \left( \proj{S_{(1,1)}}{S_{(0,2)}} + \proj{S_{(1,1)}}{S_{0,2}} + {\rm h.c.}\right).
\end{align}
Here, $J$ is the hole exchange splitting, $V_{\rm c}$ is the Coulomb energy of exciton dissociation, and
$V'_2$ is the hole energy difference between the QDs in the presence of the exciton
(taking the Coulomb correlation energy into account). The total trion Hamiltonian is 
\begin{align}
	H_{\rm T} = H_{\rm T,\u} + H_{\rm T,\d}
\end{align}
The Hamiltonian for the electron-hole short-range exchange interaction is most
conveniently written in the second quantization form, reading
\begin{align}
H^{\rm ex} &= \delta \left( a^{\dag}_\u h^{\dag}_\d h^{}_\d a^{}_\u + a^{\dag}_\d h^{\dag}_\u h^{}_\u a^{}_\d \right).
\end{align}
The considered values of the system parameters are $V_{\rm c}=20$~meV, $V_{2}=0$ (this parameter merely sets the
reference for the electric field; with this choice the $F=0$ is fixed at the hole
resonance), $V'_{2}=0.785$~meV, $\delta = 0.21$~meV, $J=0$. With these parameters, the
model is equivalent to the one used in Ref.~\cite{SchallAQT21}. \\
The dipole operators for $\sigma_+$ and  $\sigma_-$ polarized light are, respectively,
\begin{align}
\label{eq:Mplus}
M^{(+)} &= h_{1\u}a_\d = \proj{1_\d}{\d, S_{(2,0)}} + \ket{2_\d}\left(\frac{1}{\sqrt{2}}\bra{\d, S_{(1,1)}}
+ \bra{\d, T_-} + \frac{1}{\sqrt{2}} \bra{\d, T_0} \right)\,,
\end{align}    
and
\begin{align}
\label{eq:Mminus}
M^{(-)} &= h_{1\d}a_\u = \proj{1_\u}{\u, S_{(2,0)}}
 + \ket{2_\u}\left(\frac{1}{\sqrt{2}}\bra{\u, S_{(1,1)}}
+ \bra{\u, T_+} + \frac{1}{\sqrt{2}} \bra{\u, T_0} \right)\,,
\end{align}    
which determines the optically active transitions in the system.

\subsection{S2B. Calculation of 2D spectra for ultra-short pulses}
\label{sec:spectra}
~

\noindent
Using the matrix of transition amplitudes between the hole and trion states $M_{iu}$, we
can derive the two-pulse FWM spectrum in a 
general analytical way in the limit of ultra-short laser pulses. \\
The diagonalized Hamiltonian is of the form
\begin{align}
    H &= \sum_i E_{i}\left|\mathfrak{h}_i\right>\!\left<\mathfrak{h}_i\right| 
+ \sum_u E_{u}\left|\mathfrak{T}_u\right>\!\left<\mathfrak{T}_u\right| 
         + [\E(t+\tau_{12}) e^{i\varphi_1} + \E(t) e^{i\varphi_2} + {\rm h.c.} ] M\,,
\end{align}
where $\left|\mathfrak{h}_i\right>$ and $\left|\mathfrak{T}_j\right>$ are the hole and
electron eigenstates, respectively, with the corresponding eigenenergies $E_{i}$ and
$E_{u}$, 
\begin{displaymath}
\E(t) = \hat{\E}(t) e^{-i\omega t},
\end{displaymath}
and $M$ is one of the operators in Eq.~\eqref{eq:Mplus} or Eq.~\eqref{eq:Mminus}, depending
on the polarization of light. In the interaction picture this reads   
\begin{equation*}
\tilde{H}(t) = 
[\E(t+\tau_{12}) e^{i\varphi_1} + 2 \E(t) e^{i(\varphi_2+\varphi_3)} + {\rm h.c.} ] \tilde{M}(t)\,,
\end{equation*}
where
\begin{equation*}
\tilde{M}(t) = 
\sum_{i,u} M_{iu} \left| \mathfrak{T}_u\right>\!\left<\mathfrak{h}_i\right| e^{i\omega_{iu}t}\,,
\end{equation*}
with $\omega_{iu}=(E_{u}-E_{i})/\hbar$.\\
Before any optical excitation the system state is a mixture of hole
states represented by a density matrix $\rho^{(0)}$.
The first optical excitation happens at $t=-\tau_{12}$ and we assume that the laser pulse is much faster than any internal dynamics, i.e., dephasing, tunneling, or decay. The density matrix immediately after this pulse is then given by
\begin{align}
    \rho^{(1)} &= -\frac{i}{\hbar} \int\limits_{-\tau_{12}-\varepsilon}^{-\tau_{12}+\varepsilon} \left[\tilde{H}(t),\rho^{(0)}\right]\,{\rm d}t \notag\\
        &= -\frac{i}{\hbar} \int\limits_{-\infty}^{\infty} \hat{\E}^*(t+\tau_{12})
          e^{-i\varphi_1} \rho^{(0)}
           \sum_{i,u} M^*_{iu} \left| \mathfrak{h}_i\right>\!\left<\mathfrak{T}_u\right| e^{-i\omega_{iu}t}e^{-i\omega (t+\tau_{12})} \,{\rm d}t +  {\rm h.c.}\,,
\end{align}
where h.c. contains the phase $+\varphi_1$, which is irrelevant for the final FWM signal and can therefore directly be neglected. After introducing the pulse spectrum,
\begin{align}
    \hat{s}(\omega) = \int\limits_{-\infty}^{\infty} \hat{\E}(t) e^{-i\omega t}\,{\rm d}t,
\end{align}
the FWM-relevant density matrix after the first pulse reads
\begin{align}
    \rho^{(1)}(\tau_{12}) &= -\frac{i}{\hbar} e^{-i\varphi_1}\rho^{(0)}e^{i\omega_{iu}\tau_{12}} 
     \sum_{i,u} M^*_{iu} \left|
          \mathfrak{h}_i\right>\!\left<\mathfrak{T}_u\right| \hat{s}^*(\omega-\omega_{iu})
          = \sum_{iu}\rho^{(1)}_{iu} \left| \mathfrak{h}_i\right>\!\left<\mathfrak{T}_u\right|.
\end{align}
Before the interaction with the second pulse, the system can undergo dephasing and decay dynamics, such that the density matrix immediately before the next pulse is given by
\begin{align}
  \rho^{(1)}(0-\varepsilon) &= \sum_{i,u} \rho^{(1)}_{iu}(\tau_{12},\omega)
    \left| \mathfrak{h}_i\right>\!\left<\mathfrak{T}_u\right|
    e^{-\gamma_{iu}\tau_{12}}\,,
\end{align}
where $\gamma_{iu}$ is the dephasing rate for a given coherence.\\
The second pulse acts via a second order process, which leads to the density matrix
\begin{align}
    \rho^{(2)} &=  -\frac{1}{\hbar^2} \int\limits_{-\varepsilon}^{\varepsilon} \int\limits_{-\varepsilon}^{t}  \left[\tilde{H}(t),\left[ \tilde{H}(t'),\rho^{(1)}\right]\right]\,{\rm d}t'{\rm d}t \notag \\
            &= \frac{1}{\hbar^2} \int\limits_{-\varepsilon}^{\varepsilon} \int\limits_{-\varepsilon}^{\varepsilon} \tilde{H}(t)\rho^{(1)}\tilde{H}(t') \,{\rm d}t'{\rm d}t\,.
\end{align}
When considering again only contributions with the correct phase dependence for
the FWM signal ($\sim e^{2i\varphi_2}$) we arrive at 
\begin{align}
    \rho^{(2)} &= \frac{1}{\hbar^2}e^{i2\varphi_2} \int\limits_{-\infty}^{\infty}
                 \hat{\E}(t) e^{-i\omega t} \sum_{j,v} M_{jv} \left|
                 \mathfrak{T}_v\right>\!\left<\mathfrak{h}_j\right| e^{i\omega_{jv} t}
                  \rho^{(1)} \int\limits_{-\infty}^{\infty} \hat{\E}(t) e^{-i\omega t}
      \sum_{k,w} M_{kw} \left| \mathfrak{T}_w\right>\!\left<\mathfrak{h}_k\right|
      e^{i\omega_{kw} t} \notag \\ 
&= \frac{1}{\hbar^2}e^{2i\varphi_2} \sum_{j,k,v,w} \hat{s}(\omega -
              \omega_{jv})\hat{s}(\omega - \omega_{kw}) 
               M_{jv}M_{kw} \left| \mathfrak{T}_v\right>\!\left<\mathfrak{h}_j\right|
              \rho^{(1)} \left| \mathfrak{T}_w\right>\!\left<\mathfrak{h}_k\right|
              \notag\\ 
&=-\frac{i}{\hbar^2} e^{i(2\varphi_2-\varphi_1)} 
	 \sum_{i,j,k,u,v,w} \hat{s}^*(\omega - \omega_{iu})\hat{s}(\omega -
              \omega_{jv}) \hat{s}(\omega - \omega_{kw}) \notag \\ 
&\quad\times M^*_{iu}M^{ }_{jv}M^{ }_{kw} e^{(i\omega_{iu}-\gamma_{iu})\tau_{12}} 
 \left| \mathfrak{T}_v\right>\!\left<\mathfrak{h}_j\right| \rho^{(0)} 
\left| \mathfrak{h}_i\right>\!\left<\mathfrak{T}_u | \mathfrak{T}_w\right>\!\left<\mathfrak{h}_k\right| \,.
\end{align}
Finally, taking into account that the emission resulting from a coherence $\left|
  \mathfrak{T}_v\right>\!\left<\mathfrak{h}_k\right|$ is proportional to $M_{kv}^*
e^{(-i\omega_{kv}-\gamma_{kv})t}$, we arrive at the FWM signal 
\begin{align}
    \rho^{\rm (FWM)} &\sim \sum_{i,j,k,u,v}   \hat{s}^*(\omega - \omega_{iu})\hat{s}(\omega - \omega_{jv}) \hat{s}(\omega - \omega_{ku})
     \rho^{(0)}_{ij} M^*_{iu}M_{jv}M_{ku}M^*_{kv} e^{(-i\omega_{kv}-\gamma_{kv})t} e^{(i\omega_{iu}-\gamma_{iu})\tau_{12}} \,.
\end{align}
Assuming that the spectrum of the laser pulses is much broader than the energy differences
between the relevant transitions, we can set $\hat{s}(\omega-\omega_{mn}) \approx 1$. We
further assume that the initial state is a statistical mixture of the hole states, i.e.,
\begin{displaymath}
\rho^{(0)}= \sum_{n}p_{n} \left|\mathfrak{h}_n\right>\!\left<\mathfrak{h}_n\right|,
\end{displaymath} 
where we take the thermal distribution for the probabilities $p_{n}$.
With this the FWM signal can be simplified to
\begin{align}
    \rho^{\rm (FWM)}(t,\tau_{12}) &\sim \sum_n p_{n}\sum_{k,u,v}  M^*_{nu}M^{}_{nv}M^{ }_{ku}M^*_{kv} 
     e^{(-i\omega_{kv}-\gamma_{kv})t} e^{(i\omega_{nu}-\gamma_{nu})\tau_{12}}\,. \label{eq:FWM_dyn}
\end{align}
From this we directly find the 2D FWM spectrum
\begin{align}
\rho^{\rm (FWM)}(\omega,\omega_\tau) &\sim \sum_n p_{n} \sum_{k,u,v} 
f_{n u}(\omega_{\tau}-\omega_{n u}) f_{kv}^{*}(\omega - \omega_{kv}) 
 M^*_{nu}M^{ }_{nv}M^{ }_{ku}M^*_{kv} \,,\label{eq:2D_sim}
\end{align}
where $f_{iu}(\omega)$ is broadened by $\gamma_{iu}$ but also takes the spectral broadening by a
finite spectral resolution in the detection process into account. \\
\begin{figure}[h]
    \centering
    \includegraphics[width=0.5\columnwidth]{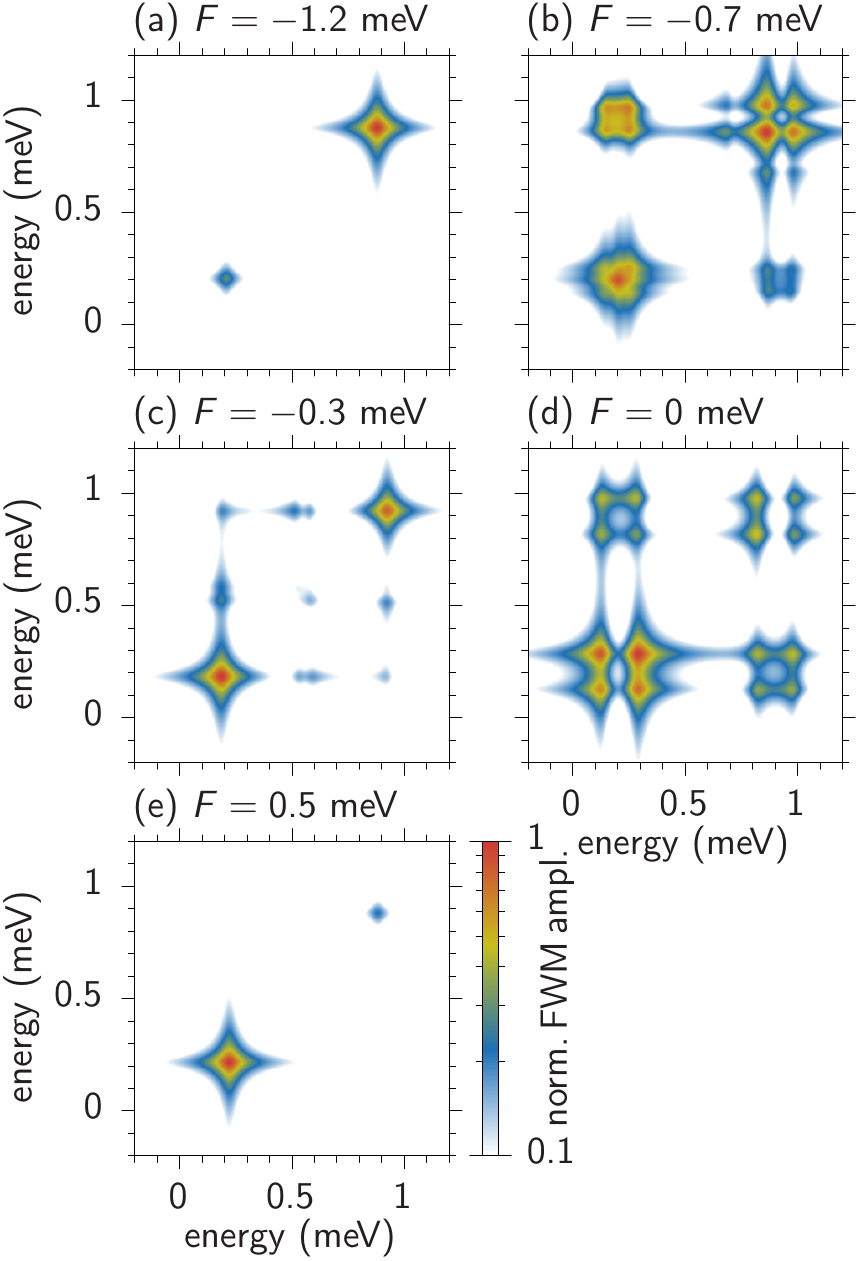}
   \caption{Simulated 2D FWM spectrum using Eq.~\eqref{eq:2D_sim} for different external electric fields as labeled in each panel.}
    \label{fig:2D_sim}
\end{figure}%
A few examples for the analytically calculated 2D spectra are shown in
Fig.~\ref{fig:2D_sim} for different bias values as given in the plot. We assume that the spectral
broadening is dominated by the instrumental resolution and use the same
value as in Fig.~2 in the main text, which roughly agrees with the
spectrometer resolution in the experiment. We clearly see that the off-diagonal peaks only
appear when the states are efficiently tunnel-coupled in the vicinity of the avoided-crossings in
the bias scan (Fig.~2 or \ref{fig:Lindblad}).

\subsection{S2C. Dynamical simulations}\label{sec:simulation}
~

\begin{figure}[h]
    \centering
    \includegraphics[width=0.5\columnwidth]{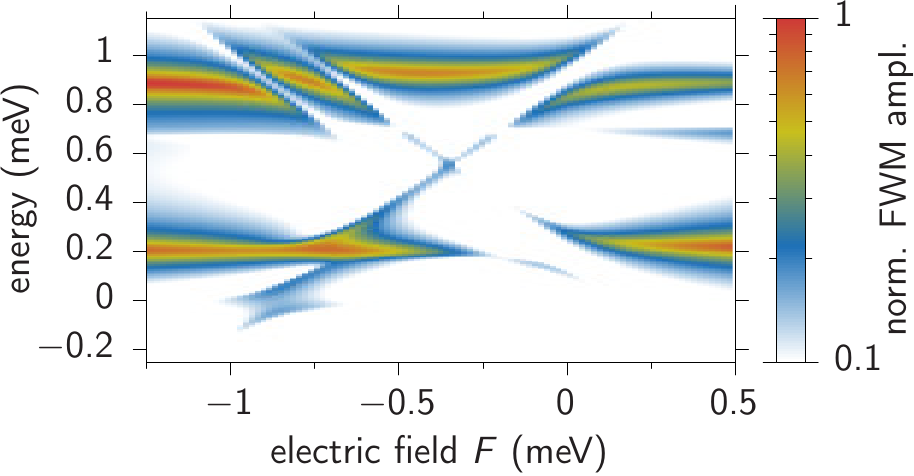}
   \caption{Theoretical bias scan of the FWM spectrum by solving the Lindblad equation and taking into account a non-vanishing pulse duration of 200~fs.}
    \label{fig:Lindblad}
\end{figure}

\noindent
To simulate the FWM signal while taking a non-vanishing pulse duration into account we calculate the system's dynamics by solving the Lindblad equation for several combinations of pulse phases $e^{i\varphi_1}$ and $e^{i\varphi_2}$~\cite{wigger2018rabi,hahn2021influence}. We then phase-filter the optically active coherences $p$ in the density matrix by numerically integrating $\int_{\varphi_1,\varphi_2}\!p e^{-i(2\varphi_2-\varphi_1)}{\rm d}\varphi_1{\rm d}\varphi_2$. For a given pulse delay $\tau_{12}$ we then determine the FWM spectrum by Fourier transforming the filtered coherence with respect to the real time $t$ after the last arriving laser pulse. The bias scan of the FWM spectrum calculated dynamically is depicted in Fig.~\ref{fig:Lindblad}. We consider a pulse duration (Gaussian standard deviation of the electric field) of 200~fs (this roughly agrees with the full width at half maximum duration of the pulse intensity of 400~fs), a pulse delay of 1~ps, and the previously determined dephasing time of 200~ps. The spectral lines are additionally broadened according to the spectral resolution of the experiment. The most important features, like the pattern of avoided-crossings and intensity distributions agree well with the analytic simulation depicted in Fig.~2 in the main text for a vanishing pulse duration. This shows that the assumption of a flat laser pulse spectrum is a valid approximation to calculate the FWM response.

\subsection{S2D. Inhomogeneous dephasing}
\label{sec:dephasing}
~

\noindent
In general, inhomogeneous dephasing of a single emitter system originates from fluctuations
of the transition energy which happens on timescales slower than a single repetition of
the FWM experiment but faster than the repetition rate of the laser. Formally, this
situation corresponds to an ensemble measurement.
In order to include such slow fluctuations, we assume the transition frequencies
$\omega_{kv},\omega_{nu}$ in Eq.~\eqref{eq:FWM_dyn} to be Gaussian random variables with
mean values $\overline{\omega}_{kv},\overline{\omega}_{nu}$ and standard deviation
$\sigma$,  which are uncorrelated if $k\neq n$ or $v\neq u$. Upon averaging over the
fluctuations, we arrive at
\begin{align*}
    \rho^{\rm (FWM)}(t,\tau_{12}) &\sim \sum_n p_{n}\sum_{k,u,v}  M^*_{nu}M^{}_{nv}M^{ }_{ku}M^*_{kv} \notag \\
     &\times e^{(-i \overline{\omega}_{kv}-\gamma_{kv})t}
     e^{(i\overline{\omega}_{nu}-\gamma_{nu})\tau_{12}}
      f_{\mathrm{deph}}(t,\tau_{12})\,. 
\end{align*}
where the dephasing factor is
\begin{equation*}
  f_{\mathrm{deph}}(t,\tau_{12}) = \left\{\begin{array}{ll}
    e^{-\sigma^2(t^2+\tau_{12}^2)}, &   k \neq n\; \mathrm{or}\; v \neq u, \\
    e^{-\sigma^2(t-\tau_{12})^{2}}, &   k = n\; \mathrm{and}\; v = u.
 \end{array} \right.
\end{equation*}
\begin{figure}[h]
    \centering
    \includegraphics[width=0.5\columnwidth]{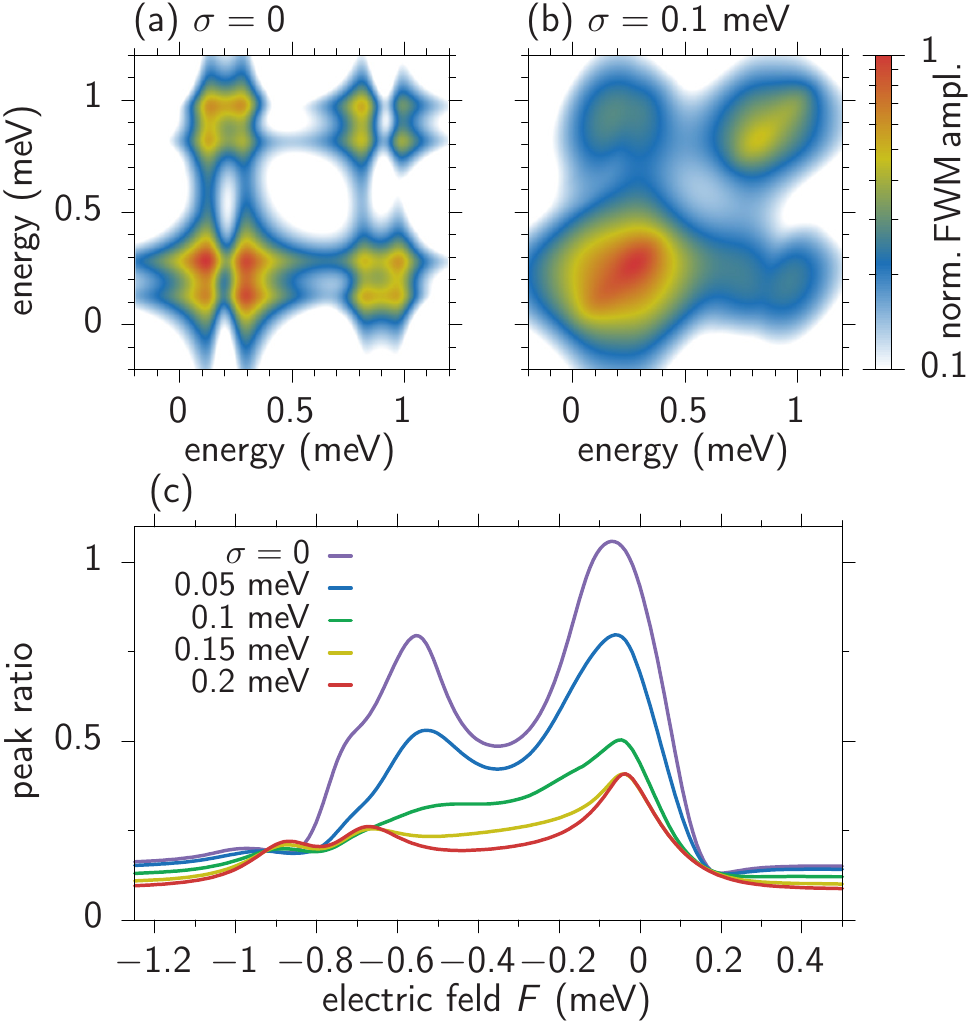}
   \caption{Impact of the inhomogeneous dephasing. (a) 2D FWM with $\sigma=0$ at an electric field of $F=0$. (b) Same as (a) but for $\sigma=0.1$~meV. (c) Simulated peak ratio curves for different inhomogeneous broadenings.}
    \label{fig:inhom}
\end{figure}%
The impact of the inhomogeneous dephasing is shown in Fig.~\ref{fig:inhom}, where we show
the 2D FWM spectrum without inhomogeneous broadening, i.e., $\sigma=0$ [$F=0$, same as
Fig.~\ref{fig:2D_sim}(d)] in (a) and $\sigma=0.1$~meV in (b).
When presenting the results, we subtract the ``background'' formed by the power-law tails
of the strong diagonal peaks in the spectral area of the off-diagonal ones. 
We clearly see that the previously sharp peaks get additionally broadened predominantly
into the diagonal direction in the 2D spectrum, as well known from several previous
works~\cite{siemens2010resonance}. In Fig.~\ref{fig:inhom}(c) we show peak ratios of the
form $\sum A_\text{off-diag}/\sum A_{\rm diag}$ as in the main text and increase the
inhomogeneous broadening as given in the plot.
The values of $A_\text{off-diag}$ and $A_{\rm diag}$ are determined by integrating
the simulated signal over the areas shown in Fig.~\ref{fig:2D}(o), with
$A_\text{off-diag}$ representing the difference of the counting results for the full
signal and for the signal with off-diagonal contributions artificially switched off (when
only the tails of the strong diagonal peaks contribute). 
As discussed in the main text, the additional dephasing significantly reduces the visibility of the off-diagonal peaks representing the coherent coupling between the different trion states.

\newpage
\subsection{S2E. k$\cdot$p model}
\label{sec:kp}
~

Our \kp model describes a vertically aligned QDM formed by InGaAs in a GaAs
matrix. Both QDs are modeled as lens-shaped with the geometry of the bottom QD being set to
18.0~nm in diameter and 4.0~nm in height, while for the top QD it is 22.0 nm and 4.4 nm,
respectively. The dots are separated by a distance of $8$~nm (base to base) and they are
placed on 0.6~nm thick wetting layers (WLs). Note, that all the values are subject to
discretization with a mesh adjusted to 0.565325~nm, which is the lattice constant of
GaAs at low temperatures. The maximum In content in the dots and in the WL is 45\% with a gradual decrease of
indium at the edges, which simulates material intermixing. This is done by processing the
composition profile with a Gaussian blur with a standard deviation of $0.6$~nm. \\
We calculate single-particle electron and hole states using an eight-band \kp model with
the envelope function approximation~\cite{bahder90,Winkler2003}. The strain which arises
due to the lattice mismatch of InAs and GaAs is determined by the continuous elasticity
approach~\cite{pryor98b}. The piezoelectric field is included up to the second order in
polarization~\cite{bester06,bester06b}, using parameters from
Ref.~\cite{Caro2015}. Details of the model are given in
Ref.~\cite{Gawarecki2018a} with further improvements of Ref.~\cite{Krzykowski2020}. The
trion states are calculated using the configuration-interaction (CI) method, where we consider a
basis of 4 electron and 8 hole single-particle states. The electric field is accounted for
at the stage of the exact diagonalization of the CI Hamiltonian~\cite{Swiderski2016}
(with the matrix elements calculated in the basis of single-particle states obtained at
zero electric field). The electron-hole exchange is also included at the CI stage. The relevant
matrix elements are calculated by mesoscopic expansion of the Coulomb interaction up to second order as in Ref.~\cite{Azizi2015}. We then estimate the atomic integrals
using rescaled hydrogen-like wave functions~\cite{Karwat2021} with the effective value of
the dielectric constant for local (on-site) integrals being equal to 4. The single- and
three-particle eigenstates are then used to compute the amplitudes of the optical
transitions~\cite{Gawelczyk2017}.\\
\begin{figure}[h]
    \centering
    \includegraphics[width=0.5\columnwidth]{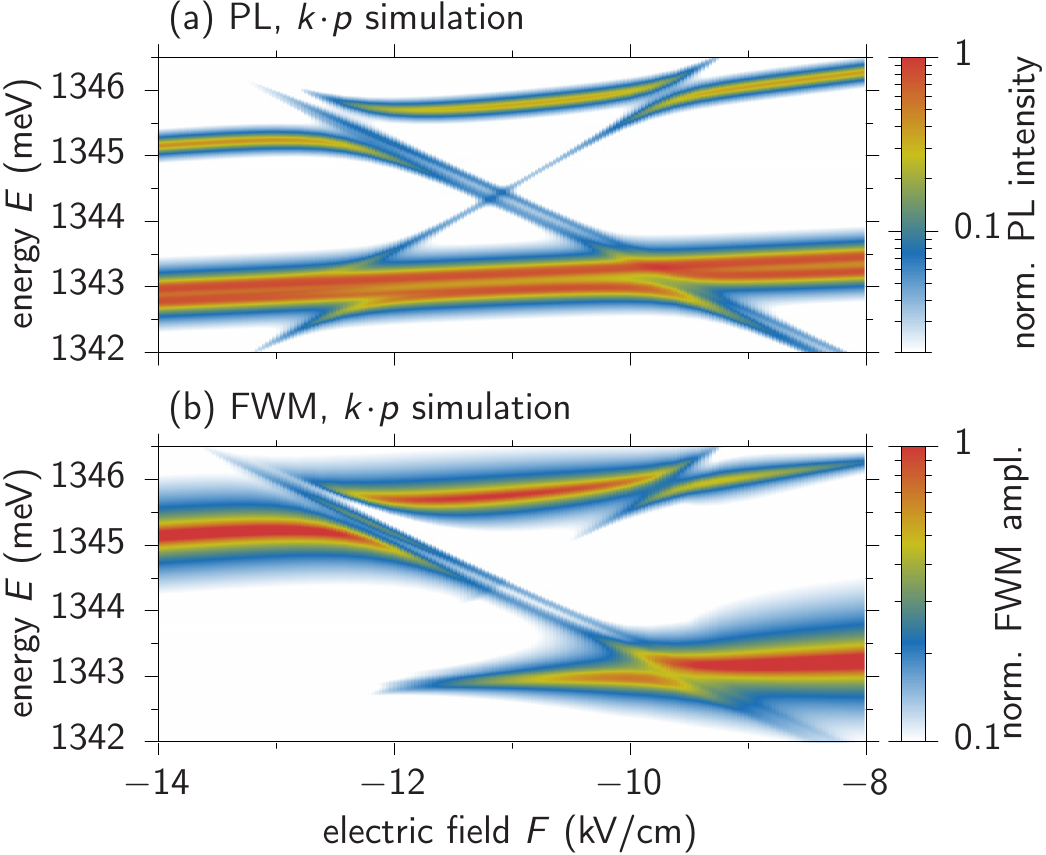}
   \caption{The luminescence (a) and FWM (b) spectra simulated using the system states obtained
     from the \kp model.}
    \label{fig:kp}
\end{figure}%
The PL and FWM spectra obtained from the \kp modelling are shown in
Fig.~\ref{fig:kp}. Qualitatively, in spite of the much reacher basis, the \kp approach
yields the same features in the PL, as the 
phenomenological model, including the hole and trion resonances as well as the
characteristic crossing of transition lines. The FWM spectrum also has exactly the same
structure, showing essentially two horizontal branches interrupted by avoided-crossings and
with intensities decaying in the opposite directions. Quantitatively, the energy scale
(which is determined by the 
Coulomb correlation energies of the three-particle system) is larger in our \kp simulation
because of a smaller inter-dot distance, which turned out to be necessary to guarantee the stability of
the relevant states in the absence of the AlGaAs barrier in our simplified geometry of the
model. One additional feature in the \kp-based PL spectrum is the line at
about 1343~meV that crosses all other transition lines without any interaction (avoided-crossing). This
line belongs to a transition between states with an excited hole, i.e., electron-hole
recombination in the presence of a spectator hole in an excited state (in the p-shell). It is visible in
our PL simulation, where all the trion states are assumed to be equally occupied. This is
consistent with the measurements shown in Fig.~\ref{fig:PL_power}, where a similar line
becomes pronounced at higher excitation powers. There is no corresponding line in the FWM
spectrum, since this is a resonantly excited coherent measurement in which only the lowest
hole states are involved.


%
\section*{References}